\documentclass[%
aip,
rsi,%
amsmath,amssymb,
reprint,%
floatfix,
]{revtex4-1}

\usepackage{graphicx}
\usepackage{dcolumn}
\usepackage{braket}
\usepackage{bm}
\usepackage{multirow}
\usepackage{csquotes}
\usepackage{amsmath}
\usepackage{amssymb}
\usepackage{multirow}

\usepackage{xcolor}

\renewcommand{\eqref}[1]{eq.~(\ref{#1})}

\begin{document}
	

\title{From Free-Energy Profiles to Activation Free Energies}

\author{Johannes C. B. Dietschreit}
\affiliation{Department of Materials Science and Engineering, Massachusetts Institute of Technology, Cambridge, Massachusetts 02139, USA}

\author{Dennis J. Diestler}
\affiliation
{University of Nebraska-Lincoln, Lincoln, Nebraska 68583, USA}

\author{Andreas Hulm}
\affiliation{Chair of Theoretical Chemistry,  Department of Chemistry, University of Munich (LMU), Butenandtstr.~7, D-81377 M\"unchen, Germany}

\author{Christian Ochsenfeld}
\affiliation{Chair of Theoretical Chemistry,  Department of Chemistry, University of Munich (LMU), Butenandtstr.~7, D-81377 M\"unchen, Germany}
\affiliation{Max Planck Institute for Solid State Research, Heisenbergstr.~1, D-70569 Stuttgart,~Germany}

\author{Rafael G\'{o}mez-Bombarelli}
\email{rafagb@mit.edu}
\affiliation{Department of Materials Science and Engineering, Massachusetts Institute of Technology, Cambridge, Massachusetts 02139, USA}

\date{20.4.2023}

		\begin{abstract}
		Given a chemical reaction going from reactant (R) to the product (P) on a potential energy surface (PES) and a collective variable (CV) discriminating between R and P, we define the free-energy profile (FEP) as the logarithm of the marginal Boltzmann distribution of the CV. 
		This FEP is not a true free energy. 
		Nevertheless, it is common to treat the FEP as the \enquote{free-energy} analog of the minimum potential energy path and to take the activation free energy, $\Delta F^\ddagger_\mathrm{RP}$, as the difference between the maximum at the transition state and the minimum at R. 
		We show that this approximation can result in large errors. 
		The FEP depends on the CV and is therefore not unique. 
		For the same reaction different, discriminating CVs can yield different $\Delta F^\ddagger_\mathrm{RP}$. 
		We derive an exact expression for the activation free energy that avoids this ambiguity.
		We find $\Delta F^\ddagger_\mathrm{RP}$ to be a combination of the probability of the system being in the reactant state, the probability density on the dividing surface, and the thermal de~Broglie wavelength associated with the transition.  
		We apply our formalism to simple analytic models and realistic chemical systems and show that the FEP-based approximation applies only at low temperatures for CVs with a small effective mass. 	    
		Most chemical reactions occur on  complex, high-dimensional PES that cannot be treated analytically and pose the added challenge of choosing a good CV. 
		We study the influence of that choice and find that, while the reaction free energy is largely unaffected, $\Delta F^\ddagger_\mathrm{RP}$ is quite sensitive.
	\end{abstract}

	\maketitle
	
	\section{Introduction}
	
	Computer simulations of chemical systems are valuable for the explanation of their experimental counterparts. 
	In the case of chemical reactions, quantities of primary interest are equilibrium constants and reaction rate constants, or quantities directly related to these, i.e., the reaction free energy $\Delta F_\mathrm{RP}$  (difference between free energies of products and reactants) and the activation free energy $\Delta F^\ddagger_\mathrm{RP}$  (the difference between free energies of transition state and reactants). 
	Indeed, the computation of such free energy differences has a long history.\cite{Kollman1993, Chipot2007, Christ2009, Chipot2014, Hansen2014, Skyner2015, Mobley2017}

	The kinetics of a chemical reaction can be modeled as a transition from a reactant well (R) on the potential energy surface (PES) to a product  well (P). 
	The two local minima are separated by a potential energy barrier that must be overcome as the atomic configuration changes and the reaction progresses. 
	The total configuration space is partitioned into (hyper) volumes corresponding to R and P by a dividing (hyper) surface, the separatrix. 
	The atomic rearrangement is described by a collective variable (CV) (or reaction coordinate), which is a function of some subset of Cartesian coordinates that gives the degree of reaction progress (e.g., 0 at R and 1 at P).
	In order to describe a reaction well, one needs to choose a \enquote{good} CV, i.e., one that distinguishes properly between configurations of R and P. 
	The CV is chosen so that it has two non-overlapping domains that correspond to the domains of R and P. 
	It is practically impossible to find the optimal CV for a complex realistic system.\cite{Bolhuis2000} 
	One must therefore base the choice of CV either on chemical intuition or on recently developed machine learning-based methods.\cite{Mendels2018a, Wang2019d, Sun2020, Bonati2020, Wang2021}
	
\begin{figure}
		\includegraphics[width=0.49\textwidth]{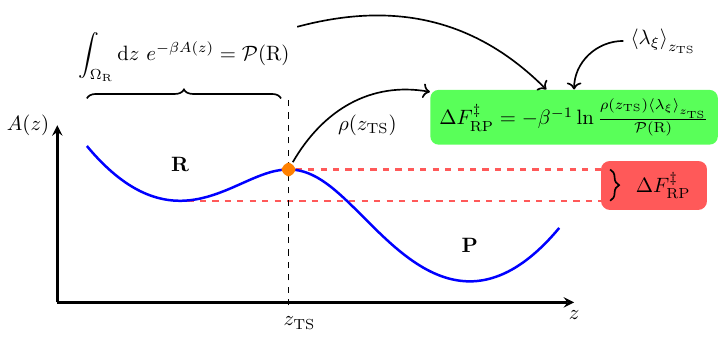}
		\caption{%
			Schematic summary of the present work showing an FEP with minima corresponding to reactant (R) and product (P) separated by a maximum. Commonly assumed, but incorrect, expression for activation free energy highlighted in red.
			The expression derived in this work is highlighted in green.  }
		\label{fig:HowTo}
	\end{figure}

	The free-energy profile (FEP)\cite{Jorgensen1989} (also referred to as the potential of mean force) is defined, up to a scaling constant, as the logarithm of the marginal Boltzmann distribution of the CV (Fig.~\ref{fig:HowTo}).
	The FEP is determined in practice by molecular dynamics (MD) or Monte Carlo simulations. 
	Because R and P are often separated by high potential energy barriers that are not overcome on simulation timescales, special simulation techniques, such as importance-sampling algorithms, must often be employed to sample configuration space properly.\cite{Valleau1977, Darve2001, Laio2002, Abrams2014, SPIWOK2015, Valsson2016, Bolhuis2002, Plotnikov2011} 
	These algorithms usually directly yield the FEP.
	
	Contrary to what the name implies, the FEP is not a true Helmholtz or Gibbs free energy.\cite{DietschreitDiestlerOchsenfeld2022}  
	Treating the FEP as if it were a free-energy analog of the minimum energy path is pervasive in the field and rarely acknowledged explicitly as the approximation that it is. 
	Differences in the FEP between local extrema are then misinterpreted as reaction and activation free energies (see red highlight in Fig.~\ref{fig:HowTo}).
	We have recently shown that this misconception leads to significant errors in reaction free energies, $\Delta F_\mathrm{RP}$.\cite{DietschreitDiestlerOchsenfeld2022} 
	The choice of the CV has a large influence on the FEP.
	In fact, the FEP has no meaning independent of the CV\cite{hartmann2007, hartmann2011two, DietschreitDiestlerOchsenfeld2022} and the structure of the FEP (e.g., the breadth and depth of local extrema or even their existence) depends on the CV. 
	Thus, a treatment that relies solely on the shape of the FEP yields CV-dependent activation free energies.
	Moreover, kinetic quantities (e.g., $\Delta F^\ddagger_\mathrm{RP}$) derived from the FEP, which depends solely on the PES and does not account for particle masses, must be approximations.
	The rigorous formula for $\Delta F^\ddagger_\mathrm{RP}$ derived here (see green highlight of Fig.~\ref{fig:HowTo} and Sec.~II~E) is independent of the precise mathematical form of the CV, as long as it discriminates between R and P. 
	We show below that a poor choice of CV has an even bigger impact on $\Delta F^\ddagger_\mathrm{RP}$ than on $\Delta F_\mathrm{RP}$.  
	
	The remainder of the article is organized as follows. 
	In Section~II we first derive an expression for the rate constant $k_\mathrm{R\rightarrow P}$. 
	Then, using the Eyring equation, we derive the connection between $\Delta F^\ddagger_\mathrm{RP}$ and $k_\mathrm{R\rightarrow P}$. 
	The physical interpretation of the components that constitute the correct activation free energy is discussed.
	In Section~III we employ simple analytic models to assess the error incurred by the common practice of taking $\Delta F^\ddagger_\mathrm{RP}$ to be the difference between the values of the FEP at the maximum (transition state) and the minimum at R. 
	Section~IV is devoted to an analysis of the sensitivity of $\Delta F_\mathrm{RP}$ and $\Delta F^\ddagger_\mathrm{RP}$ to the choice of the CV. 
	To emphasize the errors that can result from estimating $\Delta F^\ddagger_\mathrm{RP}$ directly from the FEP, we examine in Section~V a numerical one-dimensional model and two realistic chemical processes. 
	{Section~VI consists of a summary of our findings and a discussion of open questions on the computation of the activation free energy. 
		Our conclusions are summarized in Section~VII.}
	
	\section{Theory}

	\subsection{Description of the System}
	
	The interconversion of R and P is represented by the chemical reaction
	\begin{equation}
		\mathrm{R}\ \rightleftharpoons \mathrm{P} \ .
		\label{eq_reaction}
	\end{equation}
	State $\alpha$(=R,P) is defined by the region of configuration space it occupies, designated by $\Omega_\alpha$. Thus, we define the configuration integral associated with the state $\alpha$ by
	\begin{equation}
		Z_\alpha = \int_{\Omega_\alpha} \mathrm{d}\mathbf{x}\ e^{-\beta U(\mathbf{x})} \ .
		\label{eq_config_integral}
	\end{equation}
	Here $\mathbf{x}=\left(x_1, x_2, \dots, x_{3N} \right)^\mathrm{T}$ denotes the column vector of Cartesian coordinates that specify the atomic configuration; $\mathrm{d}\mathbf{x}=\prod_{i=1}^{3N}dx_i$ is the $3N$-dimensional volume element; $U(\mathbf{x})$ is the potential energy surface (PES), and $\beta \equiv 1/k_\mathrm{B}T$.
	Only those configurations $\mathbf{x}$ that belong to $\Omega_\alpha$ contribute to $Z_\alpha$, which is the effective volume of configuration space occupied by state $\alpha$.
	We assume that  $\Omega_\mathrm{R}$ and $\Omega_\mathrm{P}$  constitute the whole configuration space available to the system and they are separated by a $(3N-1)$-dimensional dividing (hyper) surface, normally taken to contain the ridge of the barrier of the PES between the minima corresponding to R and P.
	
	The course of the reaction can be monitored by a (scalar) CV (or reaction coordinate), $\xi(\mathbf{x})$, which is a function of a subset of the atomic coordinates that gives a measure of the progress of the reaction.
	The CV is chosen such that $\Omega_\mathrm{R}$ and $\Omega_\mathrm{P}$ correspond to non-overlapping domains of the CV.
	Ideally the gradient of $\xi(\mathbf{x})$ should be normal to the dividing surface, on which the CV assumes a particular value  $z_\mathrm{TS}$.
	In this case the CV discriminates properly between R and P.
	
	It is convenient to introduce mass-weighted coordinates
	\begin{equation}
		\widetilde{\mathbf{x}} = \mathbf{M}^{1/2} \mathbf{x} \ ,
		\label{eq_massweighted}
	\end{equation}
	where $\mathbf{M}$  stands for the $3N\times 3N$ diagonal matrix of atomic masses. 
	In terms of mass-weighted coordinates the Hamiltonian is
	\begin{align}
		\mathcal{H} & = \frac{1}{2}\sum_{i=1}^{3N} \widetilde{p}_i^2 + U(\widetilde{x}_1, \widetilde{x}_2, \dots, \widetilde{x}_\mathrm{3N} ) \nonumber \\
		& = \frac{1}{2}\widetilde{\mathbf{p}}^\mathrm{T}\widetilde{\mathbf{p}} + U(\widetilde{\mathbf{x}}) \ ,
		\label{eq_hamiltonian_massweighted}
	\end{align}
	where $\widetilde{p}_i = \dot{\widetilde{x}}_i$ is the momentum conjugate to the coordinate $\widetilde{x}_i$.
	Henceforth we employ the condensed notation of the second line of \eqref{eq_hamiltonian_massweighted}, where $\widetilde{\mathbf{p}}$ stands for the column vector of momenta.
	
	\subsection{Curvilinear Coordinates}
	
	The treatment of the reaction rate is facilitated by employment of a special set of coordinates, one of which is the CV.
	Hence, we transform from mass-weighted coordinates to a complete set of curvilinear coordinates, $\mathbf{q}=\mathbf{q}(\widetilde{\mathbf{x}})$, of which we take  $q_1(\widetilde{\mathbf{x}}) = \xi(\widetilde{\mathbf{x}})$.
	From the inverse transformation $\widetilde{\mathbf{x}} = \widetilde{\mathbf{x}}(\mathbf{q})$  we obtain
	\begin{equation}
		\dot{\widetilde{\mathbf{x}}} = \mathbf{J} \dot{\mathbf{q}} \ ,
		\label{eq_Jacobian_matrix}
	\end{equation}
	where $[\mathbf{J}]_{ij} = \frac{\partial \widetilde{x}_i}{\partial q_j}$ is an element of the Jacobian.
	The momentum conjugate to $\mathbf{q}$ is 
	\begin{equation}
		\mathbf{p} = \mathbf{M}_q \dot{\mathbf{q}} \ ,
		\label{eq_momentum}
	\end{equation}
	where
	\begin{equation}
		\mathbf{M}_q = \mathbf{J}^\mathrm{T}\mathbf{J} \ ,
		\label{eq_def_Mq}
	\end{equation}
	the mass matrix in curvilinear coordinates, is also referred to as the mass-metric tensor (see, for example,  Refs.~\citenum{Fixman1974, Chipot2007, denOtter2000,  denOtter2013}).
	In general, $\mathbf{M}_q$ is a full matrix.
	The Hamiltonian is given in curvilinear coordinates by
	\begin{equation}
		\mathcal{H} = \frac{1}{2} \mathbf{p}^\mathrm{T} \mathbf{M}_q^{-1} \mathbf{p} + U(\mathbf{q}) \ .
		\label{eq_Hamiltonian_curvelinear}
	\end{equation}
	From \eqref{eq_def_Mq} we deduce the following expression for the effective inverse mass matrix
	\begin{align}
		[\mathbf{M}_q^{-1}]_{ij}
		& = \sum_{k=1}^{3N} [\mathbf{J}^{-1}]_{ik} [{\mathbf{J}^{-1}}^\mathrm{T}]_{kj} \nonumber \\
		& = \left(\nabla_{\widetilde{\mathbf{x}}} q_i\right)^\mathrm{T} \left(\nabla_{\widetilde{\mathbf{x}}} q_j\right) \ ,
		\label{eq_Mq_element}
	\end{align}
	where we employ $[\mathbf{J}^{-1}]_{ik} = \frac{\partial q_i} {\partial \widetilde{x}_k}$ and $(\nabla_{\widetilde{\mathbf{x}}} q_i)^\mathrm{T} = (\partial q_i / \partial \widetilde{x}_1, \partial q_i / \partial \widetilde{x}_2, \dots, \partial q_i / \partial \widetilde{x}_{3N})$ is the $3N$-dimensional mass-weighted gradient.
	Using \eqref{eq_massweighted}, we get from \eqref{eq_Mq_element}
	\begin{equation}
		[\mathbf{M}_q^{-1}]_{ij} = (\nabla_\mathbf{x} q_i)^\mathrm{T} \mathbf{M}^{-1} (\nabla_\mathbf{x} q_j)
		\label{eq_Mq_element_cart}
	\end{equation}
	Note the distinction between $\nabla_\mathbf{x}$ for the Cartesian gradient and $\nabla_{\widetilde{\mathbf{x}}}$ for the gradient with respect to mass-weighted coordinates.

	\subsection{Reaction Rate Constant}
	
	We assume the system to be in thermodynamic equilibrium. Then the rate of the forward reaction equals the rate of the backward reaction
	\begin{equation}
		k_\mathrm{R\rightarrow P} \mathcal{P}(\mathrm{R}) = k_\mathrm{P\rightarrow R} \mathcal{P}(\mathrm{P}) \ ,
		\label{eq_kinetic_equil}
	\end{equation}
	where $k_\mathrm{R\rightarrow P}$  and $k_\mathrm{P\rightarrow R}$ are the forward and backward rate constants, and $\mathcal{P}(\mathrm{R})$  and $\mathcal{P}(\mathrm{P})$ are the respective probabilities of observing R and P.
	The rate can also be expressed in terms of the frequency $\nu$  of  crossing the dividing surface in either the forward or backward direction (i.e., of the number of times per unit time that $\xi(\widetilde{\mathbf{x}})-z_\mathrm{TS}$ changes sign).
	Since the forward and backward rates are equal, then either rate must equal  $\nu/2$.
	Thus, focusing on the forward rate, we have from \eqref{eq_kinetic_equil}
	\begin{equation}
		k_\mathrm{R\rightarrow P} = \frac{\nu}{2\mathcal{P}(\mathrm{R})}
		\label{eq_k_forward}
	\end{equation}
	The following alternative expression for the rate constant is frequently used:\cite{Berne1988, Carter1989, Hanggi1990, Hinsen1997, Bucko2017, Bailleul2020}
	\begin{equation}
		k_\mathrm{R\rightarrow P} = \frac{ \left< \dot{\xi}\ \Theta(\dot{\xi})\ \delta(\xi(\widetilde{\mathbf{x}})-z_\mathrm{TS}) \right>_{p,q}}{\left< \Theta(z_\mathrm{TS} - \xi(\widetilde{\mathbf{x}})) \right>_{p,q}}
		\label{eq_k_alternative}
	\end{equation}
	Here $\left<\ \right>_{p,q}$ denotes the ensemble average over all of phase space, $\delta$ the Dirac delta function, $\Theta$ the Heaviside function, and $\dot{\xi}$ the time derivative of the CV. 
	The equivalency of the two expressions is proven in the supplementary material.
	\\

	\subsection{Frequency of Crossing the Dividing Surface}
	
	The frequency of crossing the dividing surface can be expressed formally as the time average of the frequency with which $\xi(\widetilde{\mathbf{x}})-z_\mathrm{TS}$ changes sign\cite{Vanden2005}:
	\begin{align}
		\nu & = \lim\limits_{\tau \to \infty} \frac{1}{\tau} \int_0^{\tau} \mathrm{d}t \left| \frac{\mathrm{d}}{\mathrm{d}t} \Theta[\xi(\widetilde{\mathbf{x}}(t))-z_\mathrm{TS}]\right| \nonumber \\
		& = \lim\limits_{\tau \to \infty} \frac{1}{\tau} \int_0^{\tau} \mathrm{d}t\ \left| (\dot{\widetilde{\mathbf{x}}}(t))^\mathrm{T} \nabla_{\widetilde{\mathbf{x}}} \xi(\widetilde{\mathbf{x}}(t)) \right| \delta(\xi(\widetilde{\mathbf{x}}(t))-z_\mathrm{TS})
		\label{eq_nu_time_expr}
	\end{align}
	A proof of this expression is provided in the supplementary material.
	Assuming the system to be ergodic, we can recast the time average as an ensemble average
	\begin{equation}
		\nu = \frac{\int \mathrm{d}\widetilde{\mathbf{x}}\ \int\mathrm{d}\widetilde{\mathbf{p}}\  e^{-\beta \mathcal{H}}   \left| \dot{\widetilde{\mathbf{x}}}^\mathrm{T} \nabla_{\widetilde{\mathbf{x}}} \xi(\widetilde{\mathbf{x}}) \right| \delta(\xi(\widetilde{\mathbf{x}})-z_\mathrm{TS})}{%
			\int \mathrm{d}\widetilde{\mathbf{x}}\ \int\mathrm{d}\widetilde{\mathbf{p}}\  e^{-\beta \mathcal{H}}} \ , 
		\label{eq_nu_ens_ave}
	\end{equation}
	where $\mathcal{H}$ is given by \eqref{eq_hamiltonian_massweighted}.
	We next transform from mass-weighted to curvilinear coordinates. 
	From eqs.~(\ref{eq_Jacobian_matrix}), (\ref{eq_momentum}), and (\ref{eq_def_Mq}) we get
	\begin{equation}
		\dot{\widetilde{\mathbf{x}}}^\mathrm{T} \nabla_{\widetilde{\mathbf{x}}}\xi 
		= \mathbf{p}^\mathrm{T} \mathbf{J}^{-1} \nabla_{\widetilde{\mathbf{x}}}\xi 
		= \sum_{i=1}^{3N} p_i  (\nabla_{\widetilde{\mathbf{x}}} q_i)^\mathrm{T} \nabla_{\widetilde{\mathbf{x}}}\xi  \ ,
		\label{eq_vel_dot_gradxi}
	\end{equation}
	where the second equality invokes the definition of the inverse Jacobian. 
	Substitution of \eqref{eq_vel_dot_gradxi} into \eqref{eq_nu_ens_ave} and transformation to curvilinear coordinates yields
	\begin{widetext}
		\begin{equation}
			\nu
			= \frac{\int \mathrm{d}\mathbf{q}\ e^{-\beta U(\mathbf{q})}  \int \mathrm{d}\mathbf{p}\ e^{-\frac{\beta}{2}\mathbf{p}^T\mathbf{M}_q^{-1}\mathbf{p}}  \left| \sum_{i=1}^{3N} p_i  (\nabla_{\widetilde{\mathbf{x}}} q_i)^\mathrm{T} \nabla_{\widetilde{\mathbf{x}}}\xi \right| \delta(\xi(\mathbf{q})-z_\mathrm{TS})}{%
				\int \mathrm{d}\mathbf{q}\ e^{-\beta U(\mathbf{q})} \int \mathrm{d}\mathbf{p}\ e^{-\frac{\beta}{2}\mathbf{p}^T\mathbf{M}_q^{-1}\mathbf{p}} }
			\label{eq_nu_curvelinear}
		\end{equation}
	\end{widetext}
	To simplify this  expression we exploit the freedom afforded by curvilinear coordinates. 
	While the \enquote{first} is chosen to be the CV, the remaining $3N-1$ are as yet unspecified. 
	Hence, we require that $q_2, q_3, \dots, q_{3N}$  be \emph{orthogonal} to $q_1=\xi$, which constraint is expressed by
	\begin{equation}
		(\nabla_{\widetilde{\mathbf{x}}} q_i)^\mathrm{T} \nabla_{\widetilde{\mathbf{x}}}\xi = 0, \qquad i=2, 3, \dots, 3N
		\label{eq_orthogonality}
	\end{equation}
	In general, the construction of the orthogonal set can be achieved in a variety of ways.\cite{Darve2001}

	Invoking \eqref{eq_orthogonality}, we can express the kinetic energy as
	\begin{align}
		\frac{1}{2}\mathbf{p}^\mathrm{T}\mathbf{M}_q^{-1}\mathbf{p} &
		= \frac{1}{2} \sum_{i=1}^{3N} \sum_{j=1}^{3N} p_i (\nabla_{\widetilde{\mathbf{x}}} q_i)^\mathrm{T} (\nabla_{\widetilde{\mathbf{x}}} q_j) p_j \nonumber \\
		& = \frac{1}{2} |\nabla_{\widetilde{\mathbf{x}}} \xi|^2 p_1^2 + \sum_{i=2}^{3N} \sum_{j=2}^{3N} p_i (\nabla_{\widetilde{\mathbf{x}}} q_i)^\mathrm{T} (\nabla_{\widetilde{\mathbf{x}}} q_j) p_j \nonumber \\
		& = \frac{1}{2} |\nabla_{\widetilde{\mathbf{x}}} \xi|^2 p_1^2 + \frac{1}{2}{\mathbf{p}'}^\mathrm{T}{\mathbf{M}'}^{-1}\mathbf{p}'
		\label{eq_kinetic_energy}
	\end{align}
	where in analogy to \eqref{eq_Mq_element} we define the $(3N-1)\times(3N-1)$ inverse mass matrix ${\mathbf{M}'}^{-1}$ and the $(3N-1)$-dimensional momentum vector $\mathbf{p}' = (p_2, p_3, \dots, p_{3N})^\mathrm{T}$.
	Likewise, we can simplify \eqref{eq_vel_dot_gradxi} 
	\begin{equation}
		\sum_{i=1}^{3N} p_i  (\nabla_{\widetilde{\mathbf{x}}} q_i)^\mathrm{T} \nabla_{\widetilde{\mathbf{x}}}\xi = |\nabla_{\widetilde{\mathbf{x}}} \xi|^2 p_1
		\label{eq_dot_simple}
	\end{equation}
	Plugging eqs.~(\ref{eq_kinetic_energy}) and (\ref{eq_dot_simple}) into \eqref{eq_nu_curvelinear}, we get
	\begin{widetext}
		\begin{equation}
			\nu
			= \frac{\int \mathrm{d}\mathbf{q}\ e^{-\beta U(\mathbf{q})}\  \delta(\xi(\mathbf{q})-z_\mathrm{TS})\
				\left( \int_{-\infty}^\infty \mathrm{d}p_1\ |p_1| e^{-\frac{\beta |\nabla_{\widetilde{\mathbf{x}}} \xi|^2 p_1^2}{2}} |\nabla_{\widetilde{\mathbf{x}}} \xi|^2 \right)	
				\int \mathrm{d}\mathbf{p}'\ e^{-\frac{\beta}{2}{\mathbf{p}'}^\mathrm{T}{\mathbf{M}'}^{-1}\mathbf{p}'}  }{%
				\int \mathrm{d}\mathbf{q}\ e^{-\beta U(\mathbf{q})} \int \mathrm{d}\mathbf{p}\ e^{-\frac{\beta}{2}\mathbf{p}^\mathrm{T}\mathbf{M}_q^{-1}\mathbf{p}} }
			\label{eq_nu_orthogonality}
		\end{equation}
		Performing the integration on $p_1$ gives 
		\begin{equation}
			\nu
			= 2k_\mathrm{B}T\ \frac{\int \mathrm{d}\mathbf{q}\ e^{-\beta U(\mathbf{q})}\  \delta(\xi(\mathbf{q})-z_\mathrm{TS})\
				\cdot 1 \cdot 	
				\int \mathrm{d}\mathbf{p}'\ e^{-\frac{\beta}{2}{\mathbf{p}'}^\mathrm{T}{\mathbf{M}'}^{-1}\mathbf{p}'}  }{%
				\int \mathrm{d}\mathbf{q}\ e^{-\beta U(\mathbf{q})} \int \mathrm{d}\mathbf{p}\ e^{-\frac{\beta}{2}\mathbf{p}^\mathrm{T}\mathbf{M}_q^{-1}\mathbf{p}} }
			\label{eq_nu_integrated_p1}
		\end{equation}
		Inserting the identity $1 = |\nabla_{\widetilde{\mathbf{x}}} \xi| (2\pi k_\mathrm{B}T)^{-1/2}\int_{-\infty}^{\infty} \mathrm{d}p_1\ e^{-\frac{\beta |\nabla_{\widetilde{\mathbf{x}}} \xi|^2 p_1^2}{2}}$ into \eqref{eq_nu_integrated_p1} at the place indicated, we obtain
		\begin{equation}
			\nu
			= \sqrt{\frac{2k_\mathrm{B}T}{\pi}} 
			\frac{\int \mathrm{d}\mathbf{q}\ e^{-\beta U(\mathbf{q})}\  	
				\int \mathrm{d}\mathbf{p}\ e^{-\frac{\beta}{2}\mathbf{p}^\mathrm{T}\mathbf{M}_{q}^{-1}\mathbf{p}}\
				|\nabla_{\widetilde{\mathbf{x}}} \xi|\ \delta(\xi(\mathbf{q})-z_\mathrm{TS})}{%
				\int \mathrm{d}\mathbf{q}\ e^{-\beta U(\mathbf{q})} \int \mathrm{d}\mathbf{p}\ e^{-\frac{\beta}{2}\mathbf{p}^\mathrm{T}\mathbf{M}_q^{-1}\mathbf{p}} }
			\label{eq_nu_completed_integral}
		\end{equation}
	\end{widetext}
	Transforming back to Cartesian coordinates yields
	\begin{equation}
		\nu = \sqrt{\frac{2k_\mathrm{B}T}{\pi}} 
		\left<\delta(\xi(\mathbf{x})-z_\mathrm{TS}) \ |\nabla_{\widetilde{\mathbf{x}}} \xi|   \right> \ ,
		\label{eq_nu_solved_integrals}
	\end{equation}
	where $\left<\ \right>$ indicates the ensemble average over configuration space. 
	Using the fact that 
	\begin{equation}
		\rho(z) = \left< \delta(\xi(\mathbf{x})-z)\right> = Z^{-1} \int \mathrm{d}\mathbf{x}\ 
		\delta(\xi(\mathbf{x})-z)\ e^{-\beta U(\mathbf{x})}
		\label{eq_density_of_xi}
	\end{equation}
	is the normalized probability density of observing an atomic configuration $\mathbf{x}$ such that $\xi(\mathbf{x})=z$, we can recast \eqref{eq_nu_solved_integrals} as
	\begin{align}
		\nu & = \sqrt{\frac{2k_\mathrm{B}T}{\pi}} \rho(z_\mathrm{TS}) \left<|\nabla_{\widetilde{\mathbf{x}}} \xi| \right>_{z_\mathrm{TS}} \nonumber \\ 
		& =  \sqrt{\frac{2k_\mathrm{B}T}{\pi}} \rho(z_\mathrm{TS}) \left<\sqrt{(\nabla_{\mathbf{x}} \xi)^\mathrm{T} \mathbf{M}^{-1}(\nabla_{\mathbf{x}} \xi) } \right>_{z_\mathrm{TS}} \nonumber \\
		& = \left<\sqrt{\frac{2k_\mathrm{B}T}{\pi m_\xi}}\right>_{z_\mathrm{TS}} \rho(z_\mathrm{TS}) \ ,
		\label{eq_nu_variations}
	\end{align}
	where $\left<\ \right>_{z_\mathrm{TS}}$ signifies an average over the dividing surface. The second line of \eqref{eq_nu_variations} follows from \eqref{eq_massweighted}; the third line implicitly defines $m_\xi$, which we interpret as the effective mass of the pseudo-particle associated with the coordinate $\xi(\mathbf{x})$:
	\begin{equation}
		m_\xi^{-1} =  (\nabla_{\mathbf{x}} \xi)^\mathrm{T} \mathbf{M}^{-1}(\nabla_{\mathbf{x}} \xi) = \left[\mathbf{M}_q^{-1}\right]_{11} \ ,
		\label{eq_m_xi}
	\end{equation}
	which is the 1,1 element of the inverse mass-metric tensor (see \eqref{eq_Mq_element_cart}).\cite{Fixman1974, Chipot2007, denOtter2000, denOtter2013}
	Finally, combining eqs.~(\ref{eq_k_forward}) and (\ref{eq_nu_variations}), we obtain
	\begin{equation}
		k_\mathrm{R\rightarrow P} = \left<\sqrt{\frac{k_\mathrm{B}T}{2\pi m_\xi}}\right>_{z_\mathrm{TS}} \frac{\rho(z_\mathrm{TS}) }{\mathcal{P}(\mathrm{R})} \ .
		\label{eq_k_forward_final}
	\end{equation}
	%

	\subsection{Free Energy of Activation}
	
	Eyring’s equation relates the rate constant to a free energy of activation by defining a modified equilibrium constant for the formation of activated complex from reactant R (see, for example, Ref~\citenum{Laidler1987}). In the present notation the equation is 
	\begin{equation}
		k_\mathrm{R\rightarrow P} = \frac{k_\mathrm{B}T}{h} e^{-\beta \Delta F^\ddagger_\mathrm{RP}} \ ,
		\label{eq_Eyring}
	\end{equation}
	where $h$ is Planck's constant. 
	We use the symbol $F$ for the Helmholtz free energy in order to distinguish it from the free-energy profile denoted by $A$ (see \eqref{eq_def_fep}).
	We solve \eqref{eq_Eyring} for the activation free energy and combine the result with \eqref{eq_k_forward_final} to get 
	\begin{equation}
		\Delta F^\ddagger_\mathrm{RP} =
		- k_\mathrm{B}T \ln \frac{\rho(z_\mathrm{TS}) \left< \lambda_\xi \right>_{z_\mathrm{TS}}}{\mathcal{P}(\mathrm{R})} \ ,
		\label{eq_Addagger_accurate}
	\end{equation}
	where $\lambda_\xi \equiv \sqrt{{h^2}/{2\pi m_\xi k_\mathrm{B}T }}$. 
	We interpret $ \lambda_\xi $  as the de Broglie thermal wavelength of the pseudo-particle associated with the CV. 
	
	By expanding the logarithm in \eqref{eq_Addagger_accurate} we can recast the \enquote{exact} expression for the activation free energy as
	\begin{align}
		\Delta F^\ddagger_\mathrm{RP} & =
		- k_\mathrm{B}T \ln  \rho(z_\mathrm{TS})   + k_\mathrm{B}T \ln \mathcal{P}(\mathrm{R}) - k_\mathrm{B}T \ln \left< \lambda_\xi \right>_{z_\mathrm{TS}}  \nonumber \\
		& = A(z_\mathrm{TS}) + k_\mathrm{B}T \ln \int_{\Omega_\mathrm{R}} \mathrm{d}z\ \rho(z) - k_\mathrm{B}T \ln \left< \lambda_\xi \right>_{z_\mathrm{TS}}
		\label{eq_Addagger_expanded}
	\end{align}
	The second line of \eqref{eq_Addagger_expanded} depends on the definition of the free-energy profile (FEP)\cite{Darve2001, DietschreitDiestlerOchsenfeld2022}
	\begin{equation}
		A(z) = - k_\mathrm{B}T \ln \rho(z) \ ,
		\label{eq_def_fep}
	\end{equation} 
	and on the relation\cite{DietschreitDiestlerOchsenfeld2022}
	\begin{align}
		\mathcal{P}(\mathrm{R}) & =  \int_{\Omega_R} \mathrm{d}z\ \rho(z)  \ .
		\label{eq_AR}
	\end{align} 
	A frequently employed procedure is to set the activation free energy equal to the difference between the maximum of the FEP at $z_\mathrm{TS}$ and the minimum at $z_\mathrm{R, min}$:
	\begin{equation}
		\Delta \widetilde{F}^\ddagger_\mathrm{RP} = A(z_\mathrm{TS}) - A(z_\mathrm{R,min}) 
		\label{eq_approx_deltaA}
	\end{equation}
	We place a tilde on this formula to distinguish it from the \enquote{exact} one in \eqref{eq_Addagger_accurate}.
	Thus, $\Delta \widetilde{F}^\ddagger_\mathrm{RP}$ can be viewed as an approximation.
	For example, if the density is strongly peaked about $z_\mathrm{R, min}$, then $k_\mathrm{B}T \ln \mathcal{P}(\mathrm{R}) \approx -A(z_\mathrm{R, min})$, according to eqs.~(\ref{eq_def_fep}) and~(\ref{eq_AR}).
	Under this condition the approximate formula agrees with the exact, except for the term $-k_\mathrm{B}T \ln \left< \lambda_\xi \right>_{z_\mathrm{TS}} $.
	Therefore the influence of distortions of the coordinate system induced by $\xi(\mathbf{x})$ is ignored by $\Delta \widetilde{F}^\ddagger_\mathrm{RP}$, as is the influence of mass (see \eqref{eq_m_xi}).
	
	An alternative recasting of the exact formula for the activation free energy, \eqref{eq_Addagger_accurate}, is instructive. 
	Invoking the relations\cite{DietschreitDiestlerOchsenfeld2022}
	\begin{equation}
		q_\mathrm{R} = \frac{Z_\mathrm{R}}{\Lambda}
		\label{eq_qr}
	\end{equation}
	and
	\begin{equation}
		\mathcal{P}(\mathrm{R}) = \frac{Z_\mathrm{R}}{Z} \ ,
		\label{eq_PR_alternative}
	\end{equation}
	where $q_\mathrm{R}$ is the molecular partition function of R and $\Lambda \equiv \prod_{i=1}^{3N} \sqrt{h^2 / 2\pi m_i k_\mathrm{B}T}$ (the product of all Cartesian de Broglie wavelengths), we rewrite the exact expression as
	\begin{align}
		\Delta F^\ddagger_\mathrm{RP} & = - k_\mathrm{B}T \ln \left[ \frac{Z \rho(z_\mathrm{TS})\ \left< \lambda_\xi \right>_{z_\mathrm{TS}}}{\Lambda\ q_\mathrm{R}} \right] \nonumber \\
		& = - k_\mathrm{B}T \ln \left[ Z \rho(z_\mathrm{TS})\ \frac{\left< \lambda_\xi \right>_{z_\mathrm{TS}}}{\Lambda}  \right] +  k_\mathrm{B}T \ln q_\mathrm{R}
		\label{eq_actF_ZLambda}
	\end{align}
	The second term on the right side of \eqref{eq_actF_ZLambda} is the (negative of the) free energy of R.\cite{DietschreitDiestlerOchsenfeld2022}
	Likewise, if we regard $q^\ddagger \equiv Z \rho(z_\mathrm{TS})\ \frac{\left< \lambda_\xi \right>_{z_\mathrm{TS}}}{\Lambda} $ as the effective partition function with $z$ fixed at $z_\mathrm{TS}$, then the first term is the free energy of the constrained system.
	That $q^\ddagger$ has the stated character can be demonstrated explicitly in case the curvilinear coordinates form a complete orthogonal set. 
	Then we can rewrite \eqref{eq_actF_ZLambda} as
	\begin{align}
		\Delta F^\ddagger_\mathrm{RP} & = - k_\mathrm{B}T \ln q^\ddagger + k_\mathrm{B}T \ln q_\mathrm{R} \nonumber \\
		& = F^\ddagger - F_\mathrm{R}
		\label{eq_actF_difference}
	\end{align}
	This form of $\Delta F^\ddagger_\mathrm{RP}$ is very intuitive: 
	The activation free energy is the difference between the free energy of the system constrained to the dividing surface, $F^\ddagger$, and the free energy of the reactant, $F_\mathrm{R}$. 
	Moreover, it is noteworthy that \eqref{eq_actF_difference} assumes the same form as the corresponding expression derived by conventional transition state theory.\cite{Laidler1987}

	\section{Impact of Approximating the Activation Free Energy}
	
	In order to gauge the error incurred by approximating the activation free energy $\Delta \widetilde{F}^\ddagger_\mathrm{RP}$ (\eqref{eq_approx_deltaA}) in comparison to the \enquote{exact} $\Delta F^\ddagger_\mathrm{RP}$ (\eqref{eq_Addagger_accurate}) we study the behavior of two analytically treatable models. 
	Each consists of a single particle of mass $m$ moving in one dimension.
	The PESs are meant to represent a system with two minima, which are approximated either by square wells (SW) or parabolic (harmonic oscillator) wells (HO).
	Their detailed treatment is presented in the supplementary material.
	We take the difference between approximate and \enquote{exact} activation free energy as a correction term, which we derive to be:
	\begin{align}
		\mathrm{corr}_\mathrm{SW} = \Delta F^\ddagger_\mathrm{SW} - \Delta \widetilde{F}^\ddagger & = k_\mathrm{B}T \ln \left[ \sqrt{2\pi k_\mathrm{B}T m L_\mathrm{R}^2  / h^2} \right] 
		\label{eq_correctionSW}\\
		\mathrm{corr}_\mathrm{HO} = \Delta F^\ddagger_\mathrm{HO} - \Delta \widetilde{F}^\ddagger & = k_\mathrm{B}T \ln \left[ \sqrt{(2\pi)^2 k_\mathrm{B}^2 T^2 m  / h^2 k}\right]   \label{eq_corr_HO} 
	\end{align}
	In \eqref{eq_correctionSW} $L_\mathrm{R}$ denotes the width of the reactant square well.
	In \eqref{eq_corr_HO} $k$ is the force constant of the harmonic well.
	
	We note that $\Delta \widetilde{F}^\ddagger$ does not depend on particle mass ($m$) (as it is only derived from a marginal Boltzmann distribution), and in the one-dimensional case neither on temperature ($T$), nor parameters of the PES ($L_\mathrm{R}$ and $k$). 
	Thus, we regard the difference as a correction of $\Delta \widetilde{F}^\ddagger$ that accounts for the influence of these parameters. 
	Though the corrections for the two models exhibit different dependencies on the parameters, they can nevertheless be correlated. 
	We note directly, for example, that both corrections increase at the same rate with increasing $m$. 
	Further, both increase with increasing $T$, although $\mathrm{corr_{HO}}$ increases more rapidly. 
	Concerning the PES parameters, we observe that $\mathrm{corr_{SW}}$ increases with increasing $L_\mathrm{R}$, whereas $\mathrm{corr_{HO}}$ increases with increasing $k^{-1}$. 
	This is expected, since as $k$ decreases the harmonic potential broadens, allowing the particle to move in an effectively larger domain of R, just as an increase in $L_\mathrm{R}$ does.
	
	The one-dimensional HO model can be roughly correlated with realistic multi-dimensional systems. 
	We observe that $\nu_\mathrm{R}=\sqrt{k/m}/2 \pi$ is the frequency of oscillation of the particle about the minimum $x_\mathrm{R,min}$. 
	Hence, we can recast the correction given by \eqref{eq_corr_HO} as 
	\begin{equation}
		\mathrm{corr_{HO}} = k_\mathrm{B} T \ln(k_\mathrm{B}T / h \nu_\mathrm{R})
	\end{equation}
	For reactions carried out around room temperature $T_\circ = 300\ \mathrm{K}$, a reference frequency $\nu_\circ =  k_\mathrm{B} T_\circ /h \approx 6.0\ 10^{12}\ \mathrm{s}^{-1}$ can be defined. 
	Thus, for molecular vibrations around this frequency, the correction is negligible. 
	In the typical case, where the masses of constituent atoms (e.g., H, C, and O) are small, and the bonds are stiff, $\nu_\mathrm{R} \rightarrow \nu_\circ$ and the correction is small. 
	On the other hand, for reactions involving more massive atoms and \enquote{soft} degrees of freedom, $\nu_\mathrm{R} < \nu_\circ$ and we expect substantial corrections.
	
	\section{The Influence of the Choice of CV}
	
	The validity of the formulas describing the activation free energy (\eqref{eq_Addagger_accurate}), and the reaction free energy\cite{DietschreitDiestlerOchsenfeld2022}
	\begin{equation}
		\Delta F_\mathrm{RP} = - k_\mathrm{B}T \ln \frac{\mathcal{P}(\mathrm{P})}{\mathcal{P}(\mathrm{R})} \ ,
		\label{eq_ARP_probRandP}
	\end{equation}
	depend on the assumption that the CV distinguishes properly between R and P, as defined by the dividing surface $\mathcal{S}$. 
	Thus, knowledge of $\mathcal{S}$ is crucial to the proper choice of CV. 
	For low-dimensional model systems, the choice is generally clear, but for realistic multi-dimensional systems one usually has little or no information about $\mathcal{S}$ and must base their choice on heuristics and chemical intuition.
	Such intuitive CVs can lead to significant errors. 
	
	\begin{figure}[htb!]
		\centering
		\includegraphics[width=0.475\textwidth]{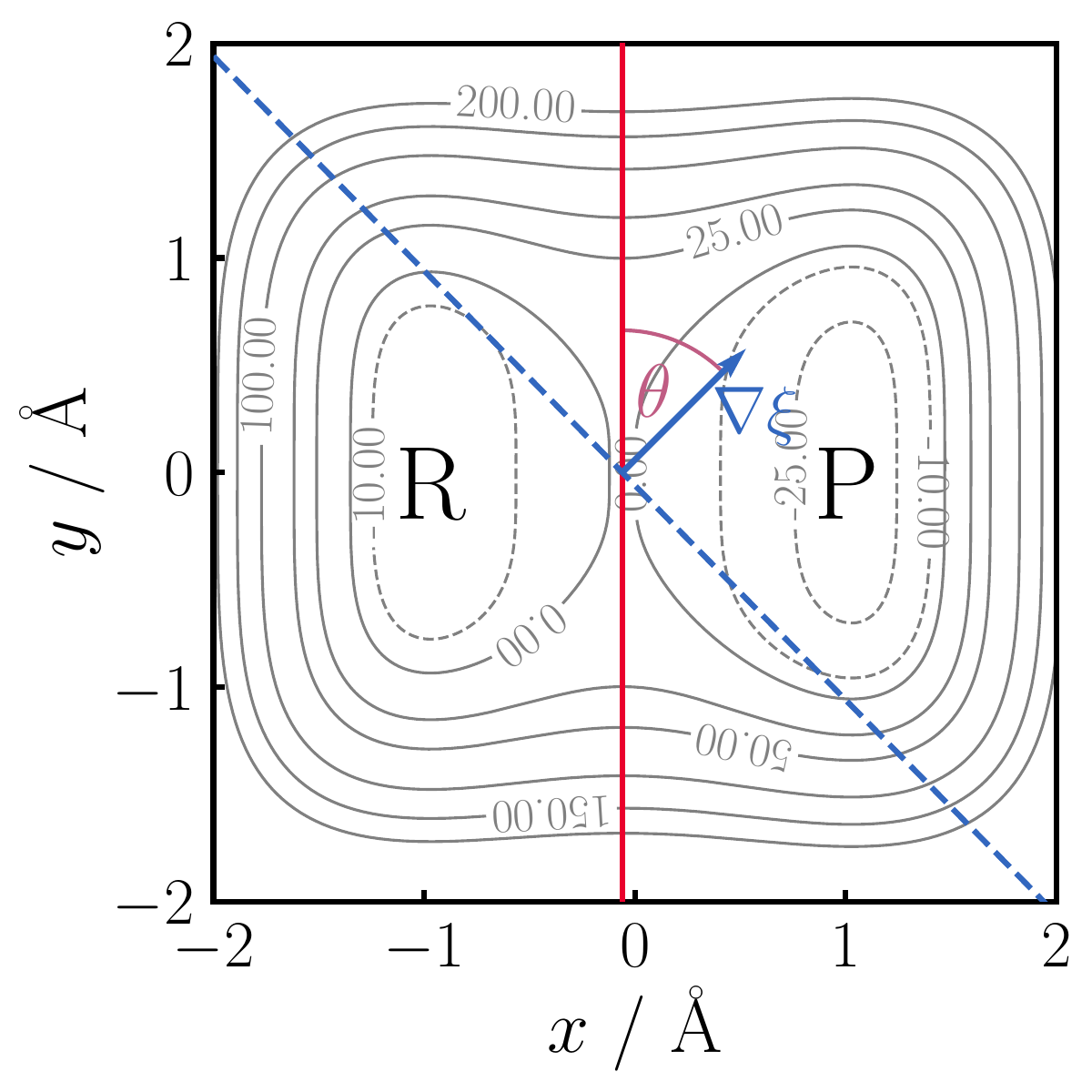}
		\caption{Contour plot of PES (\eqref{eq_pes_skwewed}) in units of kJ/mol. 
			Red line is the ideal separatrix $\mathcal{S}$. 
			Dotted blue line is the \enquote{trial} separatrix $S(\theta)$ for $\theta=45^\circ$, angle between $\nabla \xi$ (blue arrow) and $\mathcal{S}$.}
		\label{fig:CVcorrectness_pot}
	\end{figure}
	
	In this Section we systematically explore the influence of the choice of the CV on $\Delta F_\mathrm{RP}$ and $\Delta F_\mathrm{RP}^\ddagger$.
	For this purpose we employ the following model: a single particle of mass $m$ moving in two dimensions on the PES
	\begin{equation}
		U(x,y) = \epsilon \left( y^4 + x^4 - bx^2 - cx \right) \ ,
		\label{eq_pes_skwewed}
	\end{equation}
	a contour plot of which is shown in Fig.~\ref{fig:CVcorrectness_pot}. 
	The particle coordinates $x$ and $y$ are given in units of \AA\ and the energy in units of kJ/mol.
	The parameters $\epsilon$, $b$, and $c$ are taken to be 25~kJ~mol$^{-1}$~\AA$^{-4}$, 2~\AA$^{2}$ and 0.25~\AA$^{3}$, respectively. 
	The parameter $\epsilon$ effectively controls the height of the barrier of the PES between R and P; $c$ controls the difference between the minima of R and P.
	The values are chosen to yield realistic free energies ($\Delta F_\mathrm{RP}=-12.28$~kJ/mol and $\Delta F_\mathrm{RP}^\ddagger=16.06$~kJ/mol, which is roughly the activation free energy of the internal rotation of butane\cite{Murcko1996}).
	The ideal CV is $\xi(x,y) = x$ and the dividing surface coincides with the line $x = x_\mathrm{max} = -0.06725$ (see Fig.~\ref{fig:CVcorrectness_pot}).
	Clearly, $\nabla \xi \cdot \nabla U$ vanishes on $\mathcal{S}$, which is the constraint that should be obeyed by a CV that properly discriminates between R and P.\cite{Vanden2005}
	
	To vary the choice of the CV systematically, we define the CV by
	\begin{equation}
		\xi(x,y) = ax + (1-a)y
		\label{eq_xi_linearcombination}
	\end{equation}
	where $a$ is restricted to the interval $[0,1]$.
	We determine the value of $a$ by specifying the angle $\theta$ between $\nabla \xi$ and $\mathbf{e}_\mathcal{S}$, the unit vector parallel with the true separatrix $\mathcal{S}$ (i.e., $\mathbf{e}_y$).
	In other words, $a$, and therefore $\xi$, are determined by the condition $\frac{\nabla \xi}{|\nabla \xi|} \cdot \mathbf{e}_\mathcal{S} = \cos \theta$.
	(Details of the calculation are provided in the supplementary material.)
	Corresponding to a given $\theta$ (i.e., a given choice of the CV) is a \enquote{trial} separatrix $S(\theta)$, which is a line having the equation $y = -a(x-x_\mathrm{max}) / (1-a)$, where $x_\mathrm{max}$ is the $x$-coordinate of the saddle point on the PES.
	When $a=1$, then $\theta = 90^\circ$.
	In this limit $S(90^\circ)$ coincides with $\mathcal{S}$.
	As $a$ decreases from 1 to 0, $S(\theta)$ rotates counterclockwise about the point $(x_\mathrm{max}, 0)$. 
	The trial separatrix $S(45^\circ)$ is shown in Fig.~\ref{fig:CVcorrectness_pot}.
	In the limit $a=0$, $\nabla \xi = \mathbf{e}_y$, $\theta = 0$. 
	Hence, $S(0^\circ)$ is normal to $\mathcal{S}$, which makes $\xi(x,y)=y$ the worst possible choice of the CV.
	
	\begin{figure}[!tb]
		\centering
		\includegraphics[width=0.49\textwidth]{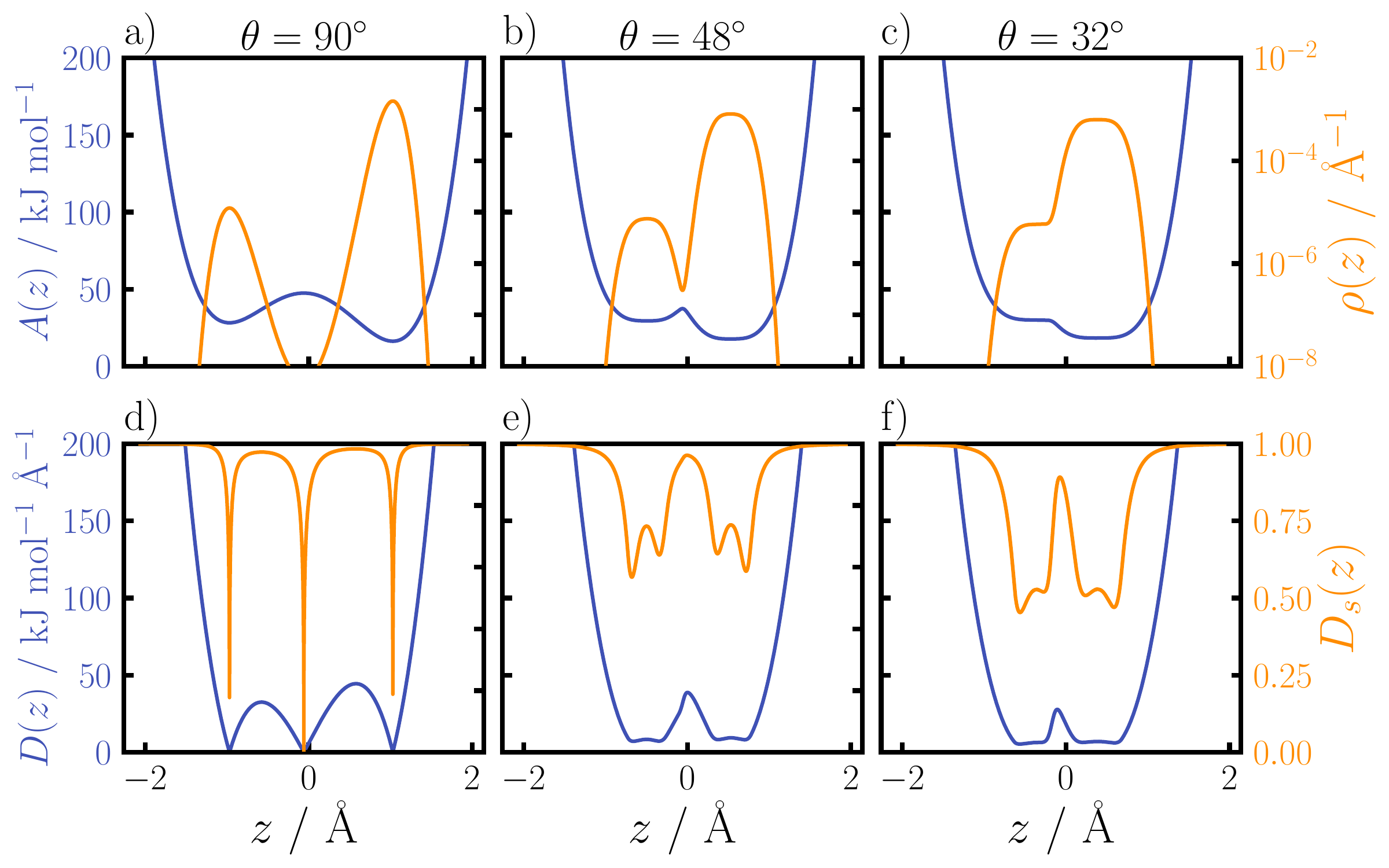}
		\caption{Top panel, plots of probability density (right ordinate, orange curve) and FEP (left ordinate, blue curve), and bottom panel $D(z)$ (left ordinate, blue curve) and $D_s(z)$ (right ordinate, orange curve) for three choices of CV:
			a) $\theta = 90^\circ$, b) $\theta = 48^\circ$ (maximum in Fig.~\ref{fig:CVcorrectness}c), and c) $\theta = 32^\circ$, last value for which the FEP still has a detectable local maximum.}
		\label{fig:CVcorrectness_three_xis}
	\end{figure}

	For a given $\theta$, we calculate the probability density $\rho(z)$ using \eqref{eq_density_of_xi}.
	As shown in the supplementary material, the evaluation of the required double integrals is facilitated by transforming from Cartesian to orthogonal coordinates $q_1 = \xi(x,y)$ and $q_2 = (a-1)x + ay$.
	We obtain the FEP using \eqref{eq_def_fep}.
	Illustrative plots of $\rho(z)$ and $A(z)$ are shown in Fig.~\ref{fig:CVcorrectness_three_xis}a-c for three CV choices.
	The local maximum of the FEP, $z_\mathrm{max}$, defines the domains of R and P.
	We note, however, that the FEPs for $\theta < 32^\circ$ lack any such local maximum.
	We henceforth ignore these choices, as the CV cannot distinguish R from P at all.
	
	As a measure of the quality of the chosen CV, we adopt a modification of the procedure introduced previously\cite{DietschreitDiestlerOchsenfeld2022}, which was to monitor the quantity $D(z) = \left<\left| \nabla \xi(\mathbf{x}) \cdot \nabla U(\mathbf{x}) \right|\right>_z$.
	We note that $D(z_\mathrm{TS})$ is exactly zero on $\mathcal{S}$ for the ideal CV (i.e., the one that discriminates perfectly between R and P).
	However, away from $\mathcal{S}$, or in case the choice of CV is not ideal, $D(z)$ is difficult to interpret, because it depends so strongly on the local gradient of the PES.
	To ameliorate this defect we propose a scaled, dimensionless orthogonality measure defined by
	\begin{align}
		D_s(z) & = \left<\left| \frac{\nabla \xi(\mathbf{x})}{|\nabla \xi(\mathbf{x})|} \cdot \frac{\nabla U(\mathbf{x})}{|\nabla U(\mathbf{x})|} \right|\right>_z \ , 
		\label{eq_Ds}
	\end{align}
	where we replace the gradients of $U$ and $\xi$ with their corresponding unit vectors.
	Thus, $D_s(z_\mathrm{TS})$ is zero on $\mathcal{S}$ for the ideal CV, where the gradients of $U$ and $\xi$ are perpendicular, and unity where they are parallel.
	
\begin{figure}[!tbh]
		\centering
		\includegraphics[width=0.475\textwidth]{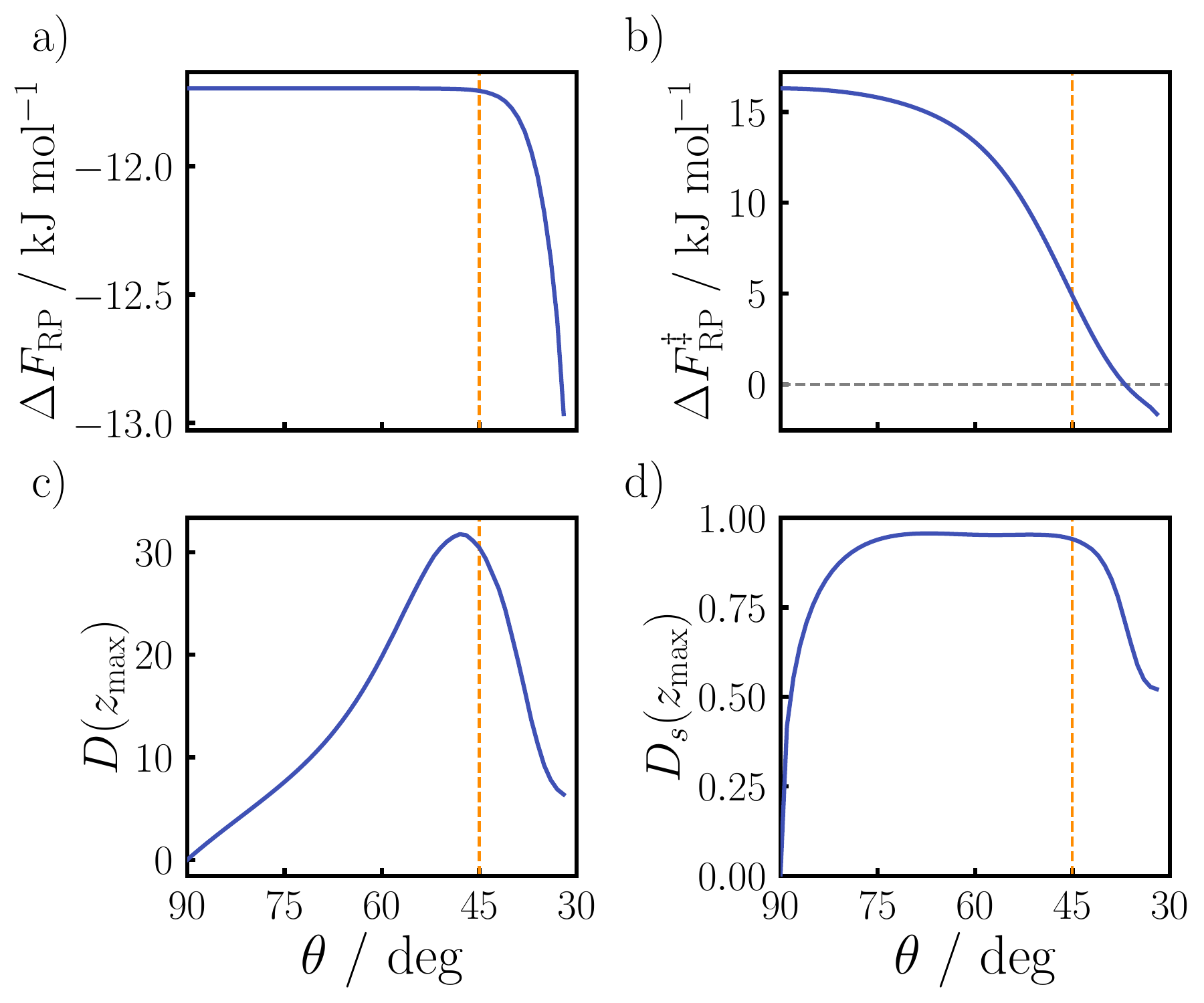}
		\caption{Plots of a) reaction free energy $\Delta F_\mathrm{RP}$, b) activation free energy $\Delta F_\mathrm{RP}^\ddagger$, c) orthogonality criterion $D(z_\mathrm{max})$, and d) scaled criterion $D_s(z_\mathrm{max})$ versus $\theta$. Orange dashed line indicates $\theta = 45^\circ$. Gray dashed line in b) guides the eye to 0~kJ/mol.}
		\label{fig:CVcorrectness}   
	\end{figure}

	One can see in Fig.~\ref{fig:CVcorrectness_three_xis}d that for the ideal CV $\xi(x,y)=x$, $D$ and $D_s$ have very sharp roots at $z_\mathrm{TS}$, indicating that the CV is orthogonal to the separatrix.
	Because of the symmetry of the PES, the two measures have two additional roots located at the minima of reactant and product.
	$D_s$ does not actually reach zero on account of the finite numerical resolution of our computation. 
	However, the sharp minima are still visible.
	Figures~\ref{fig:CVcorrectness_three_xis}e and~\ref{fig:CVcorrectness_three_xis}f show the orthogonality measure for non-ideal CVs.
	The shape of the $D$-measures changes drastically.
	Most significantly, the sharp root or minimum at the maximum of the FEP turns into a local maximum for both $D$ and $D_s$, which is an unmistakable sign that results for these CVs cannot be trusted (see the dependence of the $\Delta F^\ddagger_\mathrm{RP}$ on $\theta$ in Fig.~\ref{fig:CVcorrectness}).
	
	Using the numerically computed $\rho(z)$, we calculate the reaction free energy and activation free energy, which are given, respectively, by \eqref{eq_ARP_probRandP} and \eqref{eq_Addagger_accurate}, where we set $z_\mathrm{TS} = z_\mathrm{max}$.
	In Fig.~\ref{fig:CVcorrectness} we plot $\Delta F_\mathrm{RP}$, $\Delta F_\mathrm{RP}^\ddagger$, $D(z_\mathrm{max})$, and $D_s(z_\mathrm{max})$ as functions of $\theta$.
	Fig.~\ref{fig:CVcorrectness}d shows clearly how sensitive $D_s(z_\mathrm{max})$ is to the choice of CV.
	At $\theta = 90^\circ$ 	$D_s(z_\mathrm{max})$ vanishes, since the chosen CV coincides with the ideal one.
	But as $\theta$ decreases, $D_s(z_\mathrm{max})$ rises sharply over a narrow interval of about $10^\circ$.
	That is, for large $\theta$, $\nabla U$ and $\nabla \xi$ are almost orthogonal, whereas with decreasing $\theta$ they become nearly parallel.
	The fall off of $D_s(z_\mathrm{max})$ as $\theta$ decreases from about $45^\circ$ is due to the interference of force vectors that are almost isotropically distributed, and result in essentially randomized alignment of the force and CV gradient vectors.
	
	Since $\rho(z)$ is strongly peaked around the minima of R and P (see Fig.~\ref{fig:CVcorrectness_three_xis}), the choices of CV in the range of $45^\circ$ to $90^\circ$ separate the minima well.
	As a consequence, $\Delta F_\mathrm{RP}$ is essentially independent of the choice in this range (see Fig.~\ref{fig:CVcorrectness}a).
	In other words, over this range of choices one obtains an accurate value of the reaction free energy.
	Only for $\theta < 45^\circ$, where the CV begins to fail to discriminate between R and P, does the error in $\Delta F_\mathrm{RP}$ set in rapidly.

	As seen in Fig.~\ref{fig:CVcorrectness}b, the activation free energy is dramatically more sensitive than $\Delta F_\mathrm{RP}$ to the choice of CV.
	It deviates from the correct value by more than \enquote{chemical accuracy} (1~kcal/mol) at $\theta \approx 60^\circ$.
	For $\theta < 40^\circ$, $\Delta F_\mathrm{RP}^\ddagger$ even becomes negative.
	If this were correct, the rate of reaction would decrease with increasing temperature.
	This apparent sensitivity can be reasoned as follows.
	All points on the true separatrix have very low likelihood.
	A trial separatrix with $\theta < 90^\circ$ includes more likely configurations and therefore overestimates $\rho(z_\mathrm{TS})$.
	Since the true $\rho(z_\mathrm{TS})$ is very small, the relative error is large.
	For large probabilities, e.g., $\mathcal{P}(R)$, the same absolute error would incur a much smaller relative error.
	The relative error in the density directly translates to an absolute error in the activation free energy because of the logarithm of $\rho(z_\mathrm{TS})$ (see \eqref{eq_Addagger_expanded}).

	The fact that $\Delta F_\mathrm{RP}$ is largely unaffected by the choice of the CV explains why CVs based purely on chemical intuition can yield reaction free energies comparable with experiment. 
	However, $\Delta F_\mathrm{RP}$ is expected to become somewhat more sensitive to the choice of CV for more complex PES.
	Compared with the reaction free energy, the activation free energy is generally more sensitive.
	Hence, to achieve the same accuracy for $\Delta F_\mathrm{RP}^\ddagger$ and $\Delta F_\mathrm{RP}$ one must choose the CV with a great deal of care.

	\section{Pitfalls in the Estimation of the Activation Free Energy from the FEP}
	
	\begin{figure}[!htb]
		\centering
		\includegraphics[width=0.49\textwidth]{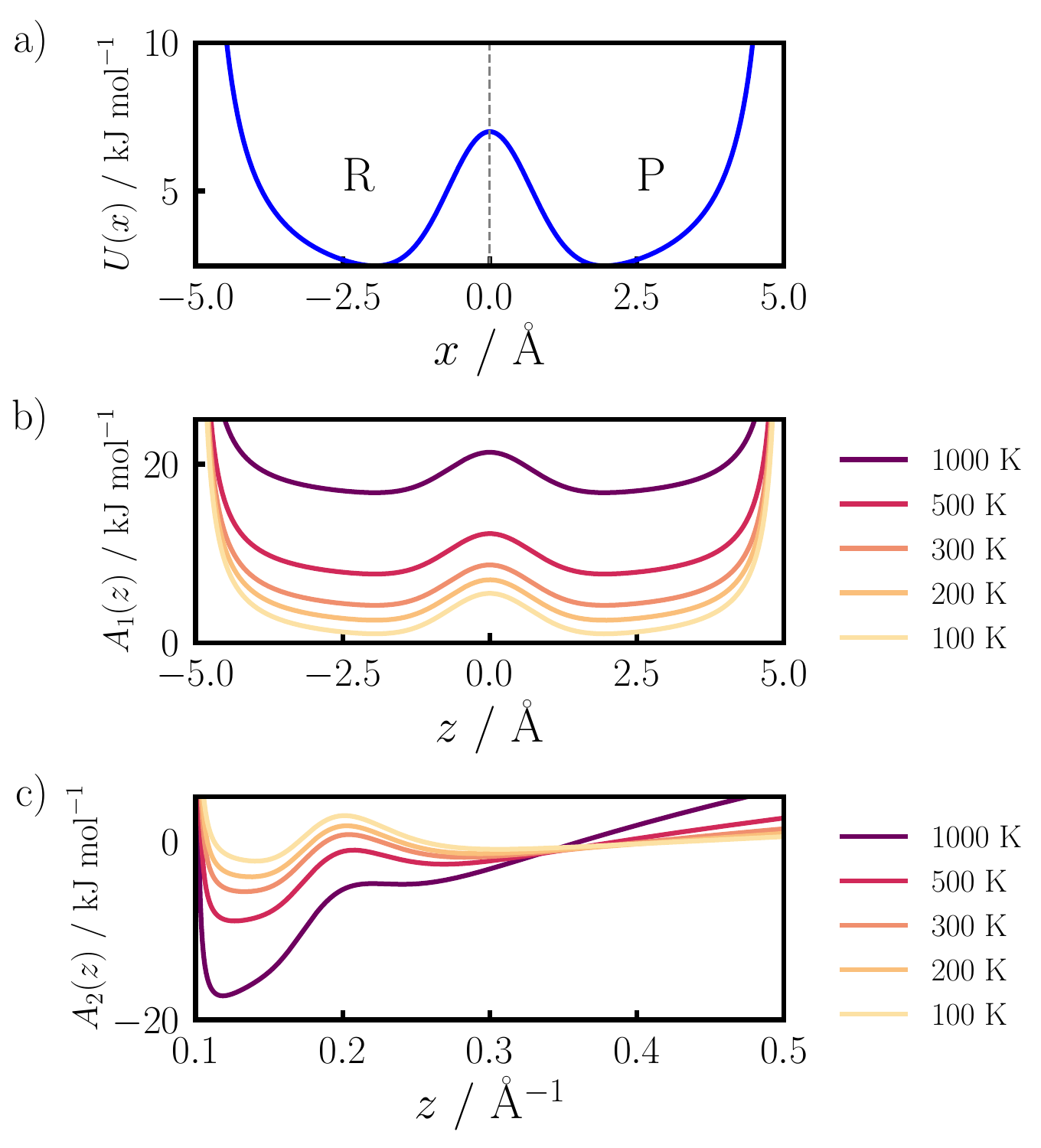}
		\caption{a) PES $U(x)$ with $\epsilon = 5$~kJ/mol (\eqref{eq:U1}); b) FEP for CV $\xi = x$ (\eqref{eq_example_A1}); c) FEP for CV $\xi = \frac{1}{x+5}$ (\eqref{eq_example_A2}).}
		\label{fig:U1}
	\end{figure}

	To further illustrate the errors that one may incur by estimating $\Delta F^\ddagger_\mathrm{RP} $  directly from the FEP alone (i.e., by invoking \eqref{eq_approx_deltaA}), we consider first a simple one-dimensional model that can be treated for the most part analytically and then models of two real chemical processes.
	
	\begin{table*}[!bt]
		\centering
		\caption{Activation free energies (kJ mol$^{-1}$) for one-dimensional model PES $U$ (\eqref{eq:U1}) with $\epsilon = 5$~kJ/mol for selections of temperatures (Kelvin) and particle masses (amu).  
			$\Delta \widetilde{F}^\ddagger_1 = A_1(z_\mathrm{TS}) - A_1(z_\mathrm{R,min})$ 
			Letters above columns specify following differences: a) $A_2(z_\mathrm{TS}) - A_2(z_\mathrm{R,min})$, b) $A_2(z_\mathrm{max}) - A_2(z_\mathrm{R,min})$, c) $A_2(z_\mathrm{TS}) - A_2(z_\mathrm{P,min})$, and d) $A_2(z_\mathrm{max}) - A_2(z_\mathrm{P,min})$.
			Numbers above columns specify particle masses.
		}
		\begin{tabular}{r| c | rrrr | rrrrr | rrrrr}
			$T$/K & $\Delta \widetilde{F}^\ddagger_1 $ & \multicolumn{4}{c|}{$\Delta \widetilde{ F}^\ddagger_2$} & \multicolumn{5}{c|}{$\Delta F^\ddagger$ (eq.~(\ref{eq_Addagger_accurate}))} & \multicolumn{5}{c}{$\Delta F^\ddagger$  (eq.~(\ref{eq_Aact_crossfreq}))$^\ast$} \\
			& & a & b & c & d &
			1 & 9 & 25 & 49 & 100 &
			1 & 9 & 25 & 49 & 100 \\\hline 
			100 & 4.53 & 3.77 & 3.78 & 5.09 & 5.10 &
			4.50 & 5.41 & 5.84 & 6.12 & 6.42 & 
			4.49$\pm$0.10 & 5.48$\pm$0.13 & 5.90$\pm$0.29 & 6.12$\pm$0.27 & 6.22$\pm$0.26\\
			200 & 4.53 & 3.10 &3.13 & 5.70 & 5.72 & 
			5.56 & 7.39 & 8.24 & 8.80 & 9.39 &
			5.57$\pm$0.09 & 7.39$\pm$0.10 & 8.24$\pm$0.11 & 8.78$\pm$0.11 & 9.29$\pm$0.17\\
			300 & 4.53 & 2.50 & 2.55& 6.35 & 6.40 & 
			6.98 & 9.72 & 11.00 & 11.84 & 12.73 &
			6.97$\pm$0.08 & 9.67$\pm$0.06 & 11.02$\pm$0.12 & 11.83$\pm$0.06 & 12.79$\pm$0.16\\
			500 & 4.53 & 1.43 & 1.58 & 7.79 & 7.94 & 
			10.39 & 14.96 & 17.08 & 18.48 & 19.97 &
			10.35$\pm$0.08 & 14.96$\pm$0.05 & 17.08$\pm$0.08 & 18.47$\pm$0.10 & 19.93$\pm$0.10\\
			1000 & 4.53 & 0.00 & 0.67& 11.91 & 12.58 & 
			20.55 & 29.69 & 33.93 & 36.73 & 39.70 & 
			20.47$\pm$0.11 & 29.63$\pm$0.16 & 33.97$\pm$0.18 & 36.74$\pm$0.19 & 39.60$\pm$0.17\\\hline
			\multicolumn{16}{l}{$^\ast \nu$ obtained from MD by means of Heaviside function (see supplementary material)}\\
			\multicolumn{16}{l}{Number after $\pm$-sign is standard deviation.}
		\end{tabular}
		\label{tab:DeltaA_U1}
	\end{table*}
	
	\subsection{One-dimensional Model}
	\label{sec:1D model}
	
	We consider a single particle of mass $m$ moving in one dimension on the PES
	\begin{equation}
		U(x) = \epsilon\  \left( \frac{b}{x+5} + e^{-ax^2} - \frac{b}{x-5} \right) \ ,
		\label{eq:U1}
	\end{equation}
	where $\epsilon$, which controls the steepness of the potential barrier, has units of kJ/mol. 
	The parameter $a$, which controls the width of the barrier, is set to 1~\AA$^{-2}$ and $b=1$~\AA.
	The PES, plotted in Fig.~\ref{fig:U1}a for the case $\epsilon = 5$~kJ/mol, has two equal minima separated by a maximum at $x=0$. 
	Because $U(x)$  diverges as $x$ approaches $-5$ or $5$, the particle is confined to the domain $-5<x<5$. 
	R and P correspond, respectively, to the domains $-5<x<0$ and $0<x<5$. 
	The symmetry of the PES dictates that $\mathcal{P}(\mathrm{R}) = \mathcal{P}(\mathrm{P})= 0.5$. 
	Therefore, from \eqref{eq_k_forward} we get
	\begin{equation}
		k_\mathrm{R\rightarrow P} = \nu / 2\mathcal{P}(\mathrm{R}) = \nu = k_\mathrm{P\rightarrow R} \ ,
		\label{eq_krp_equals_kpr}
	\end{equation}
	where $\nu$ is the crossing frequency.
	From eqs.~(\ref{eq_Eyring}) and (\ref{eq_krp_equals_kpr}), we deduce the following expression:
	\begin{equation}
		\Delta F^\ddagger_\mathrm{RP} = - k_\mathrm{B}T \ln \left(h \nu / k_\mathrm{B}T \right)
		\label{eq_Aact_crossfreq}
	\end{equation}	
	We compute $\nu$ by molecular dynamics (MD) simulation, as detailed in the supplementary material. 
	MD simulations were carried out at five temperatures in the range of 100-1000~K and for five different particle masses in the range of 1-100~amu.
	
	We consider two CVs: $\xi_1(x) = x$ and $\xi_2(x)=1/(x+5)$.
	Using \eqref{eq_def_fep}, we obtain the corresponding FEPs:
	\begin{align}
		A_1(z) & = U(z) + k_\mathrm{B}T \ln Z \label{eq_example_A1} \\
		A_2(z) & = U(z^{-1} -5) + 2k_\mathrm{B}T \ln z + k_\mathrm{B}T \ln Z \label{eq_example_A2}
	\end{align}
	Setting $\epsilon = 5$~kJ/mol ensures that even the most massive particle considered crosses the dividing surface at the lowest temperature during the 10 ns time interval of the MD simulation.
	Figs.~\ref{fig:U1}b and~\ref{fig:U1}c show plots of the FEPs based on eqs.~(\ref{eq_example_A1}) and~(\ref{eq_example_A2}). 
	We note the strong distortion of configuration space induced by $\xi_2(x)$. 
	The domains of R and P are reversed, the minima are not equal, and the maximum of the barrier between R and P does not occur precisely at $z=0.2$, the inverse of the position of the maximum of the barrier of the PES at $x=0$. 
	
	Approximate activation free energies obtained according to ~\eqref{eq_approx_deltaA} are listed in Tab.~\ref{tab:DeltaA_U1}, along with \enquote{exact} values $\Delta F^\ddagger_\mathrm{RP}$ obtained from eq.~(\ref{eq_Addagger_accurate}), which yields exactly the same result for both CVs, and from eq.~(\ref{eq_Aact_crossfreq}) via MD. 
	The excellent agreement between the values obtained from eqs.~(\ref{eq_Addagger_accurate}) and~(\ref{eq_Aact_crossfreq}) is gratifying. 
	According to \eqref{eq_example_A1}, $\Delta \widetilde{F}^\ddagger_1$ should be independent of both temperature and particle mass. 
	Likewise, $\Delta \widetilde{F}^\ddagger_2$ should depend on temperature, but we note that by definition $\Delta \widetilde{F}_2^\ddagger$ is independent of mass.
	Tab.~\ref{tab:DeltaA_U1} bears out these expectations.
	
	The dominant impression of Tab.~\ref{tab:DeltaA_U1} is the severe lack of agreement between approximate and exact activation free energies. 
	The impact of the loss of the symmetry of the PES by $\xi_2$ is particularly evident. 
	Since $z_\mathrm{TS}\approx z_\mathrm{max}$, the results in columns a and b, which correspond to the forward reaction, agree quite well, as do those of columns c and d for the backward reaction. 
	However, the magnitudes of the forward and backward activation free energies differ greatly. 
	Even more noteworthy is the contrary dependence of the activation free energy on temperature. 
	For the forward reaction it decreases with $T$, whereas for the backward reaction it increases markedly with $T$.

	\begin{table}[!bt]
		\centering
		\caption{Activation free energies (kJ mol$^{-1}$) for one-dimensional model PES $U_1$ (\eqref{eq:U1}) with $\epsilon = 50$~kJ/mol for selections of temperatures (Kelvin) and particle masses (amu).  
			$\Delta \widetilde{F}^\ddagger_1 = A_1(z_\mathrm{TS}) - A_1(z_\mathrm{R,min})$ 
			Letters above columns specify following differences: a) $A_2(z_\mathrm{max}) - A_2(z_\mathrm{R,min})$ and b) $A_2(z_\mathrm{max}) - A_2(z_\mathrm{P,min})$.
			Numbers above columns specify particle masses.}
		\begin{tabular}{r| c | rr | rrrrr }
			$T$/K & $\Delta \widetilde{F}^\ddagger_1$ & \multicolumn{2}{c|}{$\Delta \widetilde{F}^\ddagger_2$} & \multicolumn{5}{c}{$\Delta F^\ddagger$ (eq.~(\ref{eq_Addagger_accurate})} \\
			& & a & b &
			1 & 9 & 25 & 49 & 100 \\\hline 
			100 & 45.30 & 44.48 & 45.85 &
			44.32 & 45.23 & 45.66 & 45.94 & 46.23 \\
			200 & 45.30 & 43.68 & 46.40 & 
			44.50 & 46.33 & 47.18 & 47.74 & 48.33 \\
			300 & 45.30 & 42.90 & 46.96 & 
			45.12 & 47.86 & 49.14 & 49.98 & 50.87 \\
			500 & 45.30 & 41.38 & 48.09 & 
			47.12 & 51.69 & 53.81 & 55.21 & 56.70 \\
			1000 & 45.30 & 37.77 & 51.00 & 
			54.58 & 63.72 & 67.96 & 70.76 & 73.73 \\
		\end{tabular}
		\label{tab:DeltaA_U1b}
	\end{table}

	Examination of the exact data reveals the following general trends. 
	At fixed $T$, $\Delta F^\ddagger_\mathrm{RP}$ increases with particle mass $m$; the higher $T$, the greater the increase. 
	At fixed $m$, $\Delta F^\ddagger_\mathrm{RP}$ increases with $T$; the greater $m$, the greater the increase. 
	Those are the same trends observed for the analytical models in Sec.~III.
	
	To see the influence of the parameter $\epsilon$, we set $\epsilon = 50$~kJ/mol. 
	Unbiased molecular dynamics simulations were not performed for this choice of $\epsilon$ as no barrier crossings would be observed within the previously employed simulation time.
	Figure~S2 of the supplementary material displays plots of the PES and FEPs and Tab.~\ref{tab:DeltaA_U1b} lists approximate and exact free energies of activation. 
	In this case the immediate impression from Tab.~\ref{tab:DeltaA_U1b} is the greatly improved agreement between approximate and \enquote{exact} results. 
	Though the symmetry is still lost by $\xi_2$, the distortion is relatively less severe, so that forward and backward activation energies differ less. 
	The contrary dependence of forward and reverse activation energy on $T$ persists, but it is relatively weaker.   
	
	The trends in $\Delta F^\ddagger_\mathrm{RP}$ noted above for the case $\epsilon = 5$~kJ/mol hold for $\epsilon = 50$~kJ/mol, but the observed variations are relatively smaller. 
	For example, whereas the change in $\Delta F^\ddagger_\mathrm{RP}$ for $\epsilon = 5$~kJ/mol at $T=300$~K is about 80\% over the range of particle mass considered, it is only 13\% for $\epsilon = 50$~kJ/mol.
	A similar observation holds for variations of $\Delta F^\ddagger_\mathrm{RP}$ with $T$ at fixed $m$. 
	
	We stress that since both CVs perfectly distinguish between R and P, the computed \enquote{exact} activation free energy is identical for either, even though the CVs are very dissimilar.

	\subsection{Chemically Realistic Model - Mobility of Cu$^+$ in Cu-Chabazite}
	
	\begin{figure}[!hbt]
		\centering
		\includegraphics[width=0.475\textwidth]{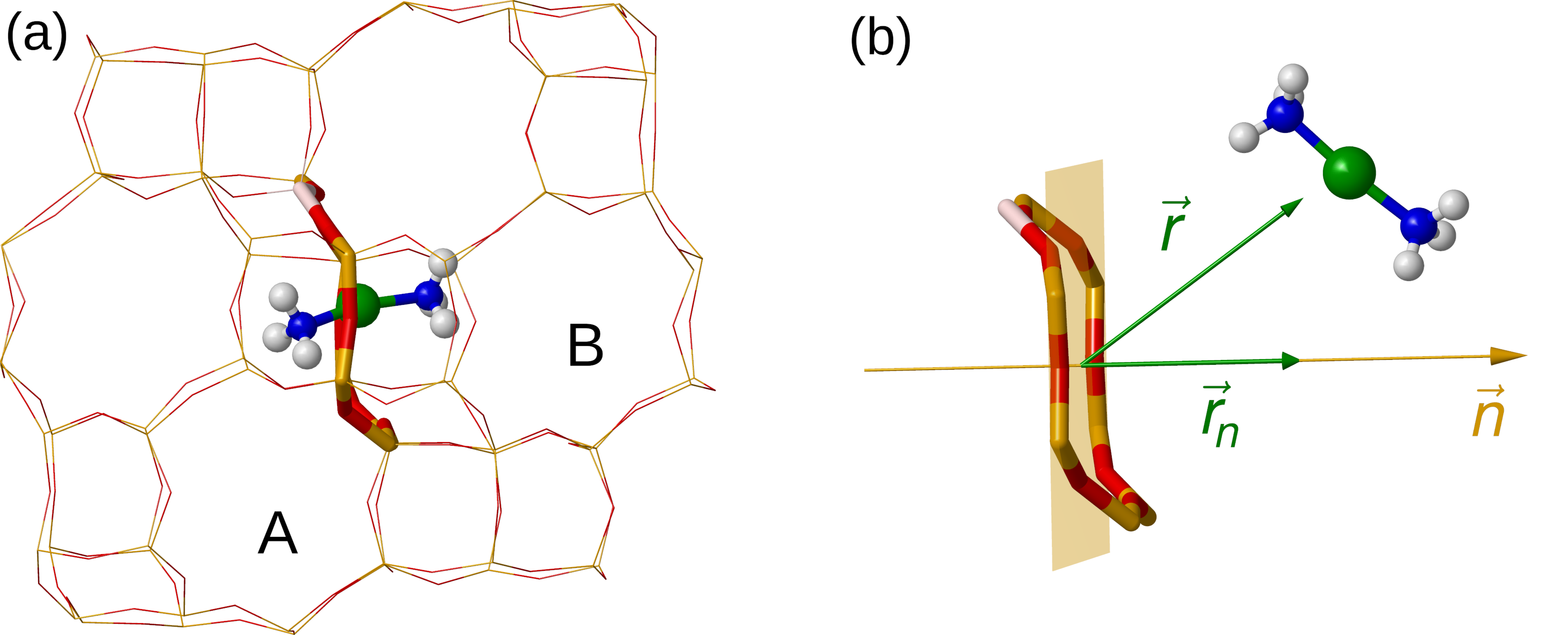}
		\caption{(a) Migration of Cu(NH$_3)_2^+$ complex from cavity A through 8-ring window into cavity B. (b) Depiction of the CV.} 
		\label{fig:SystemReisel}
	\end{figure}
	
	We consider the realistic three-dimensional model system pictured in Fig.~\ref{fig:SystemReisel}(a): 
	a [Cu(NH$_3)_2]^+$-complex migrating between cavities (A and B) in chabazite, a mixed crystal of the family of zeolites. 
	This process is of importance in the deactivation of nitrogen oxides where copper-exchanged zeolites are used as catalysts.\cite{KWAK2010187,Gao2013, Nuria2015, Borfecchia2018, PEDEN2019} 
	The migration can be regarded as a \enquote{chemical reaction}, in which the Cu-complex in cavity A or B is the \enquote{reactant} or \enquote{product}, respectively. 
	The reaction consists of the complex diffusing out of cavity A through the 8-ring (8 silicon sites) window and into cavity B. 
	Millan \textit{et al.}\cite{Millan2021} have simulated this system by means of \textit{ab initio} MD combined with umbrella sampling (for details see Ref.~\citenum{Millan2021}). 
	The CV they employ, which is depicted in Fig.~\ref{fig:SystemReisel}(b), is defined with respect to the 8-ring window that separates the cavities. 
	It is the projection of the vector position of the Cu atom onto the normal to the \enquote{average} plane of the central 4~Si and 2~O atoms of the ring that remain nearly in the same plane.
	
	Our primary purpose is to analyze the data of Millan \textit{et al.}\cite{Millan2021} in order to determine the exact values of the reaction free energy and activation free energy for the migration reaction described above. 
	We are especially interested in the effect of mass on the activation free energy. 
	The authors of Ref.~\citenum{Millan2021}  supplied the coordinates of the trajectories and the bias used for the umbrella sampling for every frame. 
	We implemented the CV in pyTorch\cite{NEURIPS2019_9015} to gain easy access to $\nabla \xi$, and consequently $m_\xi^{-1}$ (see \eqref{eq_m_xi}), through the automatic differentiation in Torch. 
	We computed the weights of every frame with an in-house implementation of MBAR.\cite{Shirts2008}
	The weights were used to re-compute the FEP and compare it with the result of Millan \textit{et al.}\cite{Millan2021}, as well as to compute the conditional ensemble average of $m_\xi^{-1}$ needed for the calculation of $\left<\lambda_\xi\right>_{z_\mathrm{TS}}$ (see \eqref{eq_Addagger_accurate}).
	
	\begin{figure}[tb]
		\centering
		\includegraphics[width=0.475\textwidth]{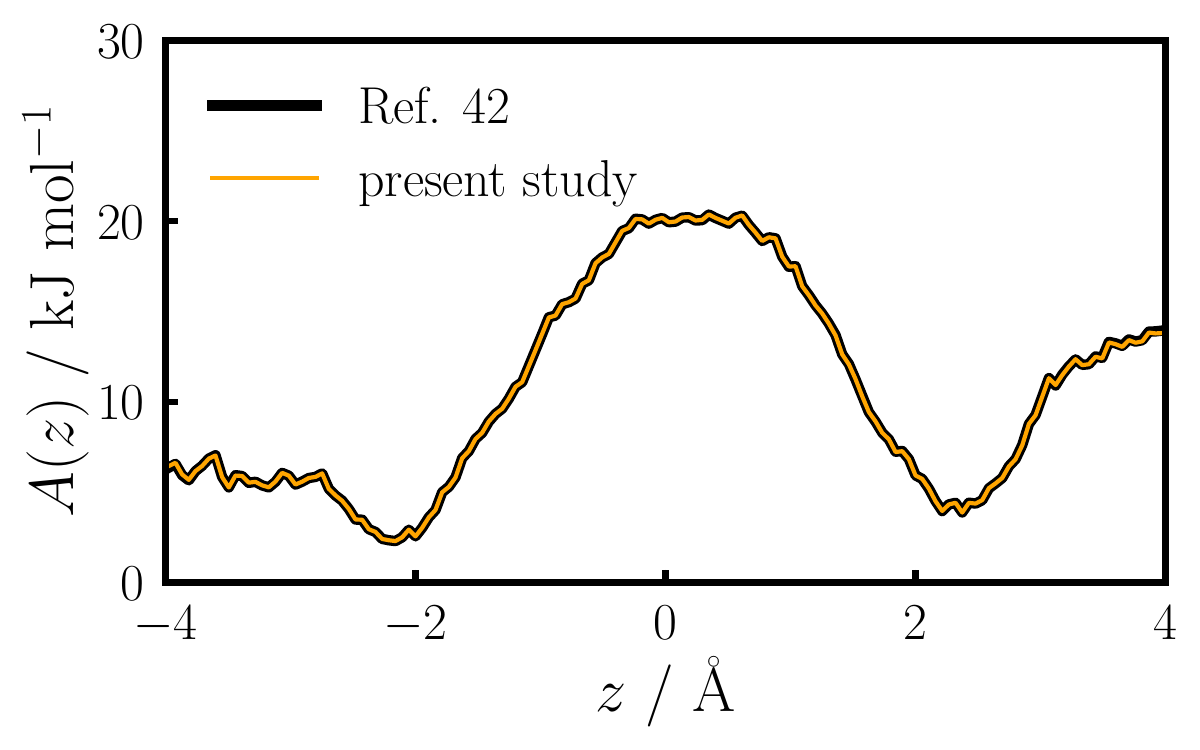}
		\caption{Comparison of FEP obtained in present study with that reported in Ref.~\citenum{Millan2021}.}
		\label{fig:ReiselPMF}
	\end{figure}
	
	The FEPs are plotted in Fig.~\ref{fig:ReiselPMF}, which shows that the agreement of our FEP with that of Millan \textit{et al.}\cite{Millan2021} is excellent. 
	The probability densities are normalized according to $\int_{-4}^{4} \mathrm{d}z\ e^{-\beta A(z)} = 1$. 
	Millan \textit{et al.}\cite{Millan2021} take the maximum of $A(z)$, located at $z = 0.35$~\AA, to be the position of the TS. 
	According to the definition of the CV, the TS should be at $z = 0.0$~\AA. 
	We computed exact and approximate reaction and activation free energies for both choices of the TS. 
	Tab.~\ref{tab:reisel} shows very clearly the large influence of mass on the activation free energy. 
	Further, the approximate free energies ($\Delta \widetilde{F}_\mathrm{AB}$ and $\Delta \widetilde{F}_\mathrm{AB}^\ddagger$) obtained by us agree well with those of Millan \textit{et al.}\cite{Millan2021}. 
	The precise choice of $z_\mathrm{TS}$ has little effect on the activation free energies, because the FEP is quite flat around $z  = 0$.
	
	Since Millan \textit{et al.}\cite{Millan2021} used the same CV for all of the systems they simulated, the correction of the activation free energy should be about the same for all. 
	Therefore the correction should not affect the ordering of the barriers ($\Delta \widetilde{F}_\mathrm{AB}^\ddagger$) they determined approximately. 
	However, we would expect any comparison with experimental activation barriers to depend strongly on the difference between the approximate and exact treatments.

	\begin{table}[!tb]
		\caption{Comparison of approximate and exact free energies (in kJ mol$^{-1}$).}
		\label{tab:reisel}
		\centering
		\begin{tabular}{p{7em} | c c c c c}
			& $z_\mathrm{TS}$/\AA & $\Delta F_\mathrm{AB}$ & $\Delta \widetilde{F}_\mathrm{AB}$ & $\Delta F_\mathrm{AB}^\ddagger$ & $\Delta \widetilde{F}_\mathrm{AB}^\ddagger$ \\\hline
			Ref.~\citenum{Millan2021} & 0.35 & -- & 1.5 & -- & 17 \\
			present study   & 0.35   & 2.8 & 1.6 & 26.2 & 18.1\\
			present study   & 0.00   & 2.8 & 1.6 & 25.8 & 17.6
		\end{tabular}
	\end{table}
	
	\subsection{Chemically Realistic Model - Radical Cyclization}
	
\begin{figure}[h!]
		\centering
		\includegraphics[width=0.75\linewidth]{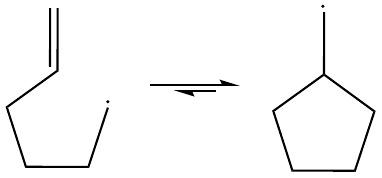}
		\caption{Scheme of the intramolecular cyclization of the reactant 5-hexenyl radical to the product methylcyclopentane radical.}
		\label{fig:cyclization_scheme}
	\end{figure}

	As a second chemical example, we consider the intramolecular cyclization of the 5-hexenyl radical (see Fig.~\ref{fig:cyclization_scheme}), a radical clock reaction.\cite{Griller1980} 
	The forward reaction involves the formation of a new single bond and the conversion of a C-C double bond to a single bond. 
	Carbon single bonds are usually stiff and have high activation barriers, as reflected in the experimental activation free energy for the cyclization, $\Delta F_\mathrm{exp}^\ddagger(300\ \mathrm{K}) = 42 \pm 4$~kJ/mol.\cite{Chatgilialoglu1991}
	Hence, we expect the approximate relation in \eqref{eq_approx_deltaA} to hold.
	As CV we choose the distance between the two carbon atoms (C1 and C5) that form a new bond, $\xi = d(\mathrm{C1-C5})$.
	The associated mass $m_\xi$ is constant  and equal to the reduced mass of the two carbon atoms (i.e., 6~amu).

	The system was simulated at 300~K by means of \textit{ab initio} MD at the $\omega$B97M-V/def2-TZVP\cite{mardirossian2015mapping, schafer1992fully} level of theory and solvated in benzene with the COSMO continuum solvation model.\cite{klamt1993cosmo}
	We employed WTM-eABF\cite{lesage2017smoothed, fu2018zooming, fu2019taming} as enhanced sampling algorithm.
	The unbiased weights were recovered with the recently developed combination of eABF and MBAR.\cite{Hulm2022}
	Details of the simulation are given in the supplementary material.

	\begin{figure}[h!]
		\centering
		\includegraphics[width=0.75\linewidth]{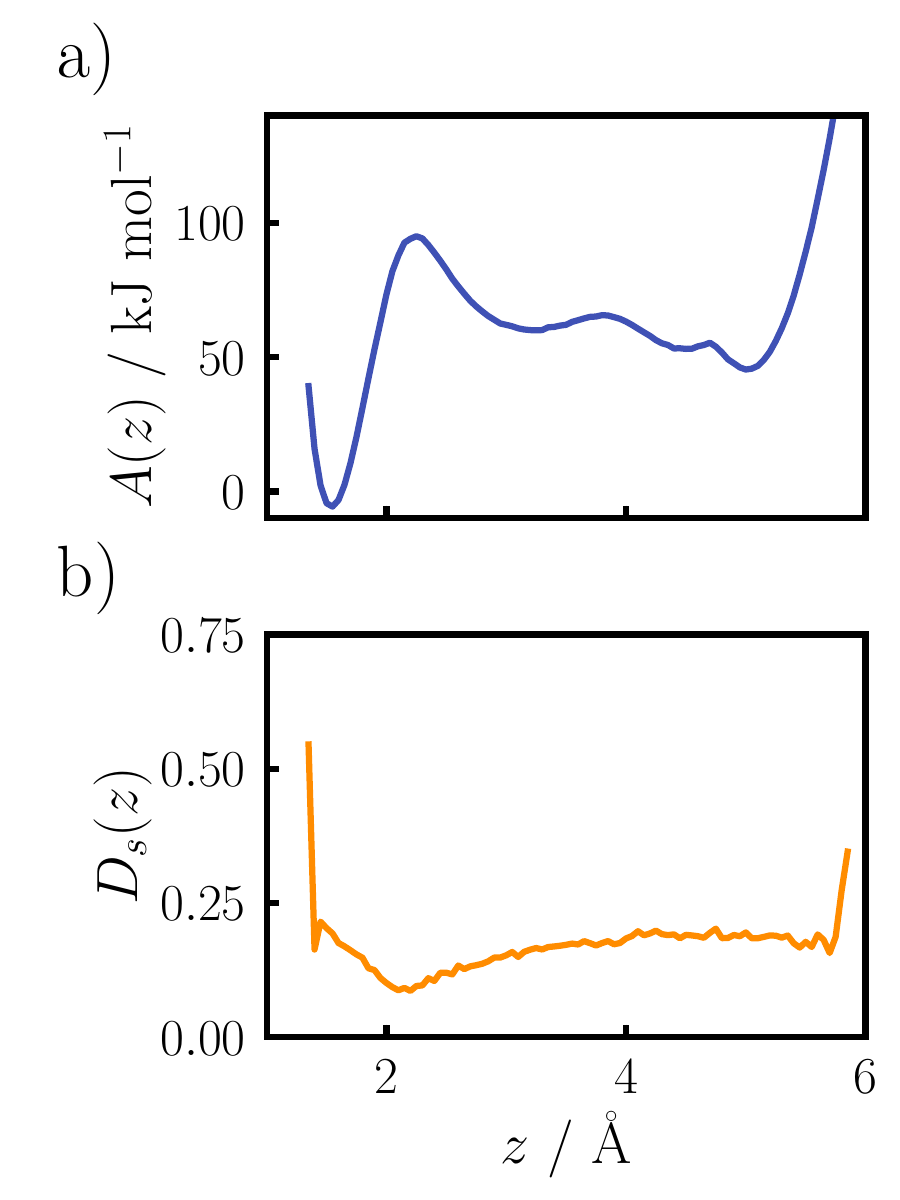}
		\caption{a) Free energy profile for the reaction shown in Fig.~\ref{fig:cyclization_scheme}.  b) Orthogonality measure $D_s(z)$.}
		\label{fig:fep_cyclization}
	\end{figure}

	The FEP (Fig.~\ref{fig:fep_cyclization}a) shows one deep minimum for P (methylcyclopentane radical) and three shallow minima for R (5-hexenyl radical).
	We take all configurations with $z > 2.2$~\AA\ to belong to R.
	The scaled orthogonality measure $D_s$ (\eqref{eq_Ds}, Fig.~\ref{fig:fep_cyclization}b) is lower than 0.25 for almost the entire range of $z$ values, rising sharply only at the ends of the simulated range.
	The plot of $D_s$ shows a clear local minimum near the local maximum of the FEP, indicating that it is a good CV.
	
	In Tab.~\ref{tab:cyclization} we can see that the exact reaction and activation free energies obtained from  eqs.~(\ref{eq_ARP_probRandP}) and~(\ref{eq_Addagger_accurate}), respectively, agree well with the approximate ones.
	Hence, this example confirms that \eqref{eq_approx_deltaA} does hold in cases of high barriers, low temperatures, light CVs, and narrow wells about the minima of R and P on the PES.
	
	\begin{table}[!h]
		\caption{Comparison of approximate and exact free energies (in kJ mol$^{-1}$) for the reaction shown in Fig.~\ref{fig:cyclization_scheme}}
		\label{tab:cyclization}
		\centering
		\begin{tabular}{c c | c c | c c}
			$\Delta {F}_\mathrm{RP}$ & $\Delta \widetilde{F}_\mathrm{RP}$ & $\Delta F_\mathrm{RP}^\ddagger$ & $\Delta \widetilde{F}_\mathrm{RP}^\ddagger$ & $\Delta F_\mathrm{PR}^\ddagger$ & $\Delta \widetilde{F}_\mathrm{PR}^\ddagger$ \\\hline
			-49.1 & -51.1 & 48.2 & 49.7 & 97.3 & 100.7\\
		\end{tabular}
	\end{table}
	
	\section{Discussion and Connection to Prior Work}
	
	This study is not the first work to present expressions for the rate constant and activation free energy based on transition state theory.\cite{Berne1988, Carter1989, Hanggi1990, Hinsen1997, Schenter2003, Bucko2017, Bailleul2020}
	However, previous work often lacks a stepwise derivation of their expression for the rate constant.
	Further, Refs.~\citenum{Schenter2003} and~\citenum{Bailleul2020}, which also present equations for the activation free energy, still include local differences of the FEP in their final expressions, which can thus be interpreted as corrections to the approximate treatment.
	Because of complex notation it is difficult to verify whether their expressions are equivalent to our \eqref{eq_Addagger_accurate}.
	It is perhaps due to the complexity of the equations and lack of physical interpretability that their expressions have not been widely adopted.
	Therefore, we are motivated to present a meticulous and straightforward derivation of the exact formula (\eqref{eq_Addagger_accurate}) for the activation free energy $\Delta F_\mathrm{RP}^\ddagger$ for the two-state process from a reactant R to a product P in a novel form. 
	The formula involves three key quantities having clear physical interpretations. 
	Two of these, $\rho(z_\mathrm{TS})$ and $\mathcal{P}(\mathrm{R}) = \int_{\Omega_\mathrm{R}} \mathrm{d}z\ \rho(z)$, depend only on $\rho(z)$, the marginal probability density that the CV $\xi(\mathbf{x})$ takes the value $z$. 
	The third, $\left<\lambda_\xi\right>_{z_\mathrm{TS}}$, can be rewritten as $\sqrt{h^2/2\pi k_\mathrm{B}T} \left<\sqrt{ m_\xi^{-1}}\right>_{z_\mathrm{TS}}$ to indicate explicitly the dependence on the effective mass of the pseudo-particle associated with the CV. 
	The three clearly defined terms also facilitate implementation.
	
	The presence of the factor $\left<\sqrt{ m_\xi^{-1}}\right>_{z_\mathrm{TS}}$ in the exact formula for $k_\mathrm{R\rightarrow P}$  (\eqref{eq_k_forward_final}) shows that knowledge of $\rho(z)$ (or alternatively $A(z)$) alone is insufficient to determine the rate constant $k_\mathrm{R\rightarrow P}$. 
	We note that in the \enquote{conventional} transition state theory\cite{Laidler1987} the rate constant is expressed in terms of canonical partition functions for reactant and activated complex (minus that associated with the CV (reaction coordinate)) and the discrete masses of the atoms enter into them. 
	In the present treatment the effective mass $m_\xi$ depends not only on the discrete masses of atoms but also on the gradient of the CV (see \eqref{eq_m_xi}). 
	If the CV is linear in the Cartesian coordinates, then $\left<\sqrt{ m_\xi^{-1}}\right>_{z_\mathrm{TS}}$  is readily expressible explicitly in terms of the discrete masses.\cite{Neria1996} 
	In general, however, the CV-conditioned ensemble average must be computed.	
	
	The \enquote{gauge-independent geometric} free-energy profile, given by
	\begin{align}
		A^G(z) & = - k_\mathrm{B} T \ln \left[ \rho(z) \left< | \nabla_{\widetilde{\mathbf{x}}} \xi | \right>_z \right] \nonumber \\
		& = - k_\mathrm{B} T \ln \left[ \rho(z) \left< \sqrt{ m_\xi^{-1}} \right>_z \right] \ .
		\label{eq:geometricA}
	\end{align}
	has been proposed\cite{hartmann2007, hartmann2011two}  as an alternative to the \enquote{standard} FEP (\eqref{eq_def_fep}). 
	Since the geometric FEP at the transition point is related to $\Delta F_\mathrm{RP}^\ddagger$ according to 
	\begin{align}
		A^G(z_\mathrm{TS}) - k_\mathrm{B} T \ln \sqrt{\frac{h^2}{2\pi k_\mathrm{B} T}} & = - k_\mathrm{B} T \ln \left[ \rho(z_\mathrm{TS}) \left< \lambda_\xi \right>_\mathrm{TS} \right] \nonumber \\
		& = \Delta F_\mathrm{RP}^\ddagger - k_\mathrm{B} T \ln \mathcal{P}(\mathrm{R}) \ ,
	\end{align}
	it is also referred to as the \enquote{kinetic} free-energy profile.\cite{Bal2020} 
	On one hand, like $A(z)$, $A^G(z)$ cannot alone provide $\Delta F_\mathrm{RP}^\ddagger$. 
	On the other, unlike $A(z)$, $A^G(z)$ cannot alone furnish $\Delta F_\mathrm{RP}$. 
	The essential reason is that $e^{-\beta A^G(z)}$ is generally not a probability density, whereas $e^{-\beta A(z)}$ always is.

	We remark on an apparent inconsistency in the dimensions of terms in \eqref{eq_Addagger_expanded}, as noted in Ref.~\citenum{Bal2020}. 
	We observe that the dimensions of $\rho(z)$ are those of $\xi^{-1}$ and the dimensions of $\left< \lambda_\xi \right>$ are those of $\xi$. 
	The argument of the logarithm is therefore dimensionless, as it should be. 
	Thus, there is no inconsistency. 
	It appears only because of the tendency to overlook that the definition of the FEP includes an implicit scaling factor, which is unfortunately rarely, if ever, pointed out. 
	The same remarks apply as well to the geometric FEP.

	\section{Conclusion}
	
	Our applications of the exact formula for the activation free energy demonstrate how significant errors can arise when $\Delta F_\mathrm{RP}^\ddagger$ is approximated simply by the difference between the values of the FEP at the transition state and reactant. 
	
	The often employed procedure to obtain $\Delta F_\mathrm{RP}^\ddagger$ solely from the FEP (by taking the difference between the values at the transition state and reactant (\eqref{eq_approx_deltaA})) is an approximation. 
	If $\rho(z)$ is strongly peaked in the vicinity of the minimum of R (i.e, at low temperature and small effective mass $m_\xi$), then \eqref{eq_approx_deltaA} may be satisfactory (see Section~V~C). 
	However, it is especially questionable when the temperature is high, $m_\xi$ is large, and the barrier of the PES between R and P is low (see Section~V~B).

	The exact formula for $\Delta F_\mathrm{RP}^\ddagger$ (\eqref{eq_Addagger_accurate}) assumes implicitly that the CV is good (i.e., it is orthogonal to the separatrix). 
	According to our study of the two-dimensional model PES with a systematically variable CV, as the CV becomes less good the reliability of $\Delta F_\mathrm{RP}^\ddagger$ decreases markedly, while that of the reaction free energy $\Delta F_\mathrm{RP}$ is only slightly affected. 
	We conclude that one must choose the CV with considerable caution in order to achieve the same accuracy for both kinetic and thermodynamic properties.

	The exact formulas for $\Delta F_\mathrm{RP}^\ddagger$ (\eqref{eq_Addagger_accurate}) and $\Delta F_\mathrm{RP}$ (\eqref{eq_ARP_probRandP}) depend only on CV-conditioned ensemble averages, which are readily available from enhanced-sampling simulations via reweighting techniques. \cite{Kumar1992, Shirts2008, tiana2008estimation,bonomi2009reconstructing, tiwary2015, schafer2020data, shirts2020statistically, Hulm2022} 
	Therefore, it should be more convenient to use these formulas than to resort to alternative special sampling strategies such as infrequent metadynamics.\cite{Dickson2017, Cossio2022} 
	
	In light of the results of the present study and those of our prior work\cite{DietschreitDiestlerOchsenfeld2022}, we recommend less reliance on the FEP alone and more on the exact formulas, which can be easily evaluated from data provided by commonly employed advanced-sampling algorithms. 
	The exact formulas are more reliable and can be clearly related to experimental data. 
	In this regard we agree with Ref.~\citenum{Bailleul2020} that use of the FEP alone should be discouraged, except we think that $\Delta F_\mathrm{RP}^\ddagger$ is a better touchstone for comparison between theory and experiment than the rate constant itself.

	\section*{Supplementary Material}
	
	The supplementary material contains the following:
	(1) proof that \eqref{eq_nu_time_expr} yields the frequency of crossing the dividing surface; (2) proof of the equivalency of eqs.~(\ref{eq_k_forward}) and~(\ref{eq_k_alternative}); (3) analytical one-dimensional models of Section III; (4) computational details of Section~IV; (5) computation of the frequency of crossing the dividing surface; (6) plots of the FEPs for models of Section~V~A with large $\epsilon$; (7) computational details of Section~V~C.

	\begin{acknowledgements}
		The authors thank Dr.~Reisel Millan, who provided full access to their simulations of chabazite and furnished Fig.~6.
		J.C.B.D. is thankful for the support of the Leopoldina Fellowship Program, German National Academy of Sciences Leopoldina, grant number LPDS 2021-08.
		C.O. acknowledges financial support  by the \enquote{Deutsche Forschungsgemeinschaft} (DFG, German Research Foundation) within cluster of excellence \enquote{e-conversion} (EXC 2089/1-390776260) and SFB 1309-325871075 \enquote{Chemical Biology of Epigenetic Modifications} and further support as Max-Planck-Fellow at the MPI-FKF Stuttgart.
		R.G.-B. acknowledges support from the Jeffrey Cheah Career Development Chair.
	\end{acknowledgements}
	
	\section*{Author Declarations}
	\subsection*{Conflict of Interest}
	The authors have no conflicts of interest to disclose.

	\section*{Data Availability Statement}
	
	The data that support the findings of this study are available from the corresponding author upon reasonable request.
	
	\section*{References}
	\bibliography{references}

	\clearpage
	
	\appendix

	\begin{center}
		\LARGE
		Supporting Material for: From Free-Energy Profiles to Activation Free Energies
	\end{center}
	
	\section{Proof that Eq.~(14) Yields the Frequency of Crossing the Dividing Surface}
	
	Starting with eq.~(14) of the article,
	\begin{equation}
		\nu = \lim\limits_{\tau \to \infty} \frac{1}{\tau} \int_0^{\tau} \mathrm{d}t \left| \frac{\mathrm{d}}{\mathrm{d}t} \Theta[\xi(\mathbf{x}(t))-z_\mathrm{TS}]\right| \ ,
		\label{eq:nu_heaviside}
	\end{equation}
	we  apply the chain rule of differentiation to obtain
	\begin{align}
		\nu & = \lim\limits_{\tau \to \infty} \frac{1}{\tau} \int_0^{\tau} \mathrm{d}t \left| \frac{\mathrm{d} \Theta}{\mathrm{d}\xi} \dot{\xi}(t) \right| \nonumber \\
		& =  \lim\limits_{\tau \to \infty} \frac{1}{\tau} \int_0^{\tau} \mathrm{d}t \left| \frac{\mathrm{d} \Theta}{\mathrm{d}\xi} \right| \left| \dot{\xi}(t) \right| \nonumber \\
		& = \lim\limits_{\tau \to \infty} \frac{1}{\tau} \int_0^{\tau} \mathrm{d}t \left| \delta[\xi(t) - z_\mathrm{TS}] \right| \left| \dot{\xi}(t) \right|
		\label{eq:nu_prod_absolutes}
	\end{align}
	We now utilize the property of the Dirac distribution \cite{Cohen1977}
	\begin{equation}
		\delta[f(t)] = \sum_i \frac{\delta(t-t_i)}{\left| (\mathrm{d}f / \mathrm{d}t)_{t=t_i} \right| } \ ,
	\end{equation}
	where $f(t_i)=0$ and $(\mathrm{d}f / \mathrm{d}t)_{t=t_i} \neq 0$, to recast \eqref{eq:nu_prod_absolutes} as
	\begin{align}
		\nu & \lim\limits_{\tau \to \infty} \frac{1}{\tau} \int_0^{\tau} \mathrm{d}t \left( \sum_i \frac{\delta(t-t_i)}{\left|\dot{\xi}(t_i)\right| } \left|\dot{\xi}(t)\right| \right) \nonumber \\
		& = \lim\limits_{\tau \to \infty} \frac{1}{\tau} \sum_i \frac{1}{\left|\dot{\xi}(t_i)\right|} \int_0^{\tau} \mathrm{d}t\ \delta(t-t_i) \left|\dot{\xi}(t)\right| \nonumber \\
		& = \lim\limits_{\tau \to \infty} \frac{1}{\tau}  \sum_{j=1}^{N_\tau} 1\nonumber \\
		& = \lim\limits_{\tau \to \infty} \frac{N_\tau}{\tau} 
	\end{align}
	where $N_\tau$  is the number of zeroes of $\xi(t) - z_\mathrm{TS}$ on the interval  $[0, \tau]$, which is equal to the number of times $\xi(t) - z_\mathrm{TS}$ changes sign during the interval. 
	Hence, $N_\tau/\tau$ is just the frequency of crossing the dividing surface.
	
	\section{Proof of the Equivalency of  Eqs.~(12) and~(13)}
	
	We assume that in general the CV is \enquote{good} in that it distinguishes properly between R and P (i.e., no configuration of R has the same value of the CV as a configuration of P).
	Moreover, implicit in eq.~(13) is the assumption that the value of $z_\mathrm{TS} -\xi(\mathbf{x})$ is positive for configurations of R and negative for those of P.
	It follows that
	\begin{equation}
		\left< \Theta[z_\mathrm{TS} - \xi(\widetilde{\mathbf{x}})]\right>_{p,q} = \mathcal{P}(\mathrm{R}) ,
	\end{equation}
	where the Heaviside function is 1 in the domain of R, where its argument is positive.

	Assuming the system to be ergodic, we can replace the time average in eq.~(S2) with the ensemble average
	\begin{equation}
		\nu = \left< \delta[\xi(\widetilde{\mathbf{x}}) - z_\mathrm{TS}] \left| \dot{\xi}(\widetilde{\mathbf{x}})\right| \right>_{p,q} \ .
	\end{equation}
	We note that since $\nu$ is the frequency of crossings due to both forward and reverse reactions, the absolute value of the rate of change of the CV is necessary to prevent cancellations of forward and reverse contributions. 
	Further, because the system is taken to be in thermodynamic equilibrium, the forward and reverse reactions occur with the same frequency.
	Hence, we can simply count reactions in one direction, say forward from R to P, where $\dot{\xi} > 0$.
	Then the absolute value of $\dot{\xi}$ becomes unnecessary. 
	The sign of the velocity is enforced by introducing a Heaviside function.
	Thus, we have
	\begin{equation}
		\nu / 2 = \left< \delta[\xi(\widetilde{\mathbf{x}}) - z_\mathrm{TS}]\ \dot{\xi}(\widetilde{\mathbf{x}})\ \Theta(\dot{\xi}) \right>_{p,q}
	\end{equation}
	This is precisely the numerator of the expression in eq.~(13) in the article.
	Dividing eq.~(S7) by~(S5) finally gives
	\begin{equation}
		\frac{\nu}{2\mathcal{P}(\mathrm{R})} = \frac{\left< \delta[\xi(\widetilde{\mathbf{x}}) - z_\mathrm{TS}]\ \dot{\xi}(\widetilde{\mathbf{x}})\ \Theta(\dot{\xi}) \right>_{p,q}}{\left< \Theta[z_\mathrm{TS} - \xi(\widetilde{\mathbf{x}})]\right>_{p,q}}
	\end{equation}

	\section{Analytical One-Dimensional Models of Section~III}
	
	Here we treat a one-dimensional system consisting of a single particle of mass $m$ moving on a PES with two minima separated by a maximum. 
	We consider two model PESs, comparing the approximate and \enquote{exact} free energies of activation.
	
	We take the PES of the first to be a square well specified piecewise by
	\begin{equation}
		U_\mathrm{SW}(x) = \left\{ 
		\begin{array}{r l }
			\infty, & x < 0 \\
			\epsilon_\mathrm{R}, & 0 < x < L_\mathrm{R}-\delta \\
			\epsilon_\mathrm{B}, & L_\mathrm{R}-\delta < x < L_\mathrm{R}+\delta \\
			\epsilon_\mathrm{P}, & L_\mathrm{R}+\delta < 0 < L \\
			\infty, & L < x
		\end{array}
		\right.
	\end{equation}
	Assuming that $\delta \ll L$, $\epsilon_\mathrm{B} > \epsilon_\mathrm{R}$, and $\epsilon_\mathrm{B} > \epsilon_\mathrm{P}$, and taking the CV to be $\xi(x) = x$, we derive the probability density
	\begin{equation}
		\rho_\mathrm{SW}(z) = Z^{-1} e^{-\beta U_\mathrm{SW}(z)} \ .
	\end{equation}
	Hence the probability of observing R is
	\begin{equation}
		\mathcal{P}_\mathrm{SW}(\mathrm{R}) = \int_{\Omega_\mathrm{R}} \mathrm{d}z\ \rho_\mathrm{SW}(z) = Z^{-1} L_\mathrm{R}e^{-\beta \epsilon_\mathrm{R}}
	\end{equation}	
	According to eq.~(30), we have for the \enquote{exact} free energy of activation
	\begin{align}
		\Delta F^\ddagger_\mathrm{SW} & = -k_\mathrm{B}T \ln \left[ \frac{\rho_\mathrm{SW}(z_\mathrm{TS}) \left<\lambda_\xi\right>_{z_\mathrm{TS}} }{\mathcal{P}_\mathrm{SW}(\mathrm{R})} \right] \nonumber \\
		& = -k_\mathrm{B}T \ln \left[ \frac{e^{-\beta \epsilon_\mathrm{B}}}{Z} \frac{Z}{L_\mathrm{R} e^{-\beta \epsilon_\mathrm{R}}} \left<\lambda_\xi\right>_{z_\mathrm{TS}} \right] \nonumber \\
		& = \epsilon_\mathrm{B} -  \epsilon_\mathrm{R} -k_\mathrm{B}T \ln \left[ \sqrt{h^2 / 2\pi m k_\mathrm{B}T L_\mathrm{R}^2} \right]
		\label{eq_DeltaAddagger_SW}
		\ ,
	\end{align}
	where we use the relation $\left<\lambda_\xi\right>_{z_\mathrm{TS}} = \sqrt{h^2/2\pi m k_\mathrm{B}T}$.
	
	For the second model we consider a double-well PES having minima of $\epsilon_\mathrm{R}$ at $x_\mathrm{R,min}$ and $\epsilon_\mathrm{P}$ at $x_\mathrm{P,min}$, separated by a maximum of $\epsilon_\mathrm{B}$ at the transition state. 
	We approximate this PES about the minima by the harmonic-oscillator (HO) approximation (e.g., $U(x) \approx U_\mathrm{HO}(x) = \epsilon_\mathrm{R} + k/2(x-x_\mathrm{R,min})^2$, where the force constant is $k = \left. \frac{\mathrm{d}^2 U}{\mathrm{d}x^2}\right|_{x=x_\mathrm{R,min}}$). 
	We again take the CV to be $\xi(x)=x$. 
	Thus, the probability of observing R is 
	\begin{align}
		\mathcal{P}(\mathrm{R}) & = \int_{\Omega_\mathrm{R}} \mathrm{d}z\ e^{-\beta U(z)} Z^{-1} \nonumber \\
		\approx Z^{-1} \int_{-\infty}^{\infty} \mathrm{d}x\ e^{-\beta \left(\epsilon_\mathrm{R} + \frac{k}{2}(x - x_\mathrm{R,min})^2 \right)} \nonumber \\
		= \frac{e^{-\beta \epsilon_\mathrm{R}}}{Z} \sqrt{\frac{2\pi}{k \beta}} \ ,
		\label{eq_SWapprox_1Dexmaple}
	\end{align}
	where we approximate the probability density in the domain of R by $\rho_\mathrm{HO}(\xi) =Z^{-1} e^{-\beta U_\mathrm{HO}(x)}$.
	Using eq.~(30), we obtain the \enquote{exact} activation free energy 
	\begin{align}
		\Delta F^\ddagger_\mathrm{HO} & = -k_\mathrm{B}T \ln \left[ \frac{\rho_\mathrm{HO}(z_\mathrm{TS}) \left<\lambda_\xi\right>_{z_\mathrm{TS}} }{\mathcal{P}_\mathrm{HO}(\mathrm{R})} \right] \nonumber \\
		& = -k_\mathrm{B}T \ln \left[ \frac{e^{-\beta \epsilon_\mathrm{B}}}{Z} \frac{Z}{e^{-\beta \epsilon_\mathrm{R}}} \sqrt{\frac{k}{2\pi k_\mathrm{B}T}} \sqrt{h^2 / 2\pi m k_\mathrm{B}T}  \right] \nonumber \\
		& = \epsilon_\mathrm{B} -  \epsilon_\mathrm{R} -k_\mathrm{B}T \ln \sqrt{h^2 k /[(2\pi k_\mathrm{B})^2 m T^2]} 
		\label{eq_HOapprox_1Dexmaple}
	\end{align}
	where we again invoke the relation $<\lambda_\xi>_{z_\mathrm{TS}} =\sqrt{h^2/2\pi m k_\mathrm{B}T}$.
	
	According to eq.~(34), the approximate activation free energy for both models is given by:
	\begin{align}
		\Delta \widetilde{F}^\ddagger & = A(z_\mathrm{TS}) - A(z_\mathrm{R, min}) = k_\mathrm{B} T\ \ln \frac{\rho(z_\mathrm{TS})}{\rho(z_\mathrm{R, min})}  \nonumber \\
		& = U(z_\mathrm{TS}) - U(z_\mathrm{R, min}) =
		\epsilon_\mathrm{B} - \epsilon_\mathrm{R}
		\label{eq_approxA_1D}
	\end{align}
	for both models. Comparing \eqref{eq_approxA_1D} with \eqref{eq_DeltaAddagger_SW} and with \eqref{eq_HOapprox_1Dexmaple}, we see that the difference between \enquote{exact} and approximate activation free energies is, respectively
	\begin{align}
		\mathrm{corr_{SW}} = \Delta F^\ddagger_\mathrm{SW} - \Delta \widetilde{F}^\ddagger & = k_\mathrm{B}T \ln \left[ \sqrt{2\pi k_\mathrm{B}T m L_\mathrm{R}^2  / h^2} \right] 
		\\
		\mathrm{corr_{HO}} = \Delta F^\ddagger_\mathrm{HO} - \Delta \widetilde{F}^\ddagger & = k_\mathrm{B}T \ln \left[ \sqrt{(2\pi)^2 k_\mathrm{B}^2 T^2 m  / h^2 k}\right] \ . 
	\end{align}
	%

	\section{Computational Details of Section~IV}
	
	Required numerical computations are handled by NumPy \cite{harris2020array}.
	Plots are generated with Matplotlib \cite{Hunter:2007}.
	
	\subsection{Determination of the Parameter $a$}
	
	The CV in Section~IV of the article is given by eq.~(40)
	\begin{equation}
		\xi(x,y) = ax + (1-a)y \ ,
		\label{eq:cv_sec_iv}
	\end{equation}
	where $a$ is restricted to the interval $[0,1]$. 
	It is determined by specifying the angle $\theta$ between
	$\nabla \xi$ and $\mathbf{e}_\mathcal{S}$, the unit vector parallel with the true separatrix $\mathcal{S}$ (i.e., $\mathbf{e}_y$). 
	The angle is related to the two vectors by
	\begin{equation}
		\cos \theta = \frac{\nabla \xi}{|\nabla \xi|} \cdot \mathbf{e}_\mathcal{S} \ ,
		\label{eq:cos_theta}
	\end{equation}
	where $\theta$ is restricted to the interval $[0, \pi / 2]$.
	From \eqref{eq:cv_sec_iv} we obtain
	\begin{equation}
		\nabla \xi = a \mathbf{e}_x + (1-a)\mathbf{e}_y \ .
		\label{eq:grad_xi_sec_iv}
	\end{equation}
	Substitution of \eqref{eq:grad_xi_sec_iv} into \eqref{eq:cos_theta} yields
	\begin{equation}
		\cos \theta = \frac{a \mathbf{e}_x + (1-a)\mathbf{e}_y}{\sqrt{a^2 + (1-a)^2}} \cdot \mathbf{e}_y = \frac{ (1-a)}{\sqrt{a^2 + (1-a)^2}} \ .
		\label{eq:cos_theta_of_a}
	\end{equation}
	Solving this equation for $a$, we get
	\begin{equation}
		a_{\pm} = \frac{\sin^2 \theta}{sin^2 \theta - \cos^2 \theta} \pm \sqrt{\frac{\sin^4 \theta}{(sin^2 \theta - \cos^2 \theta)^2} - \frac{\sin^2 \theta}{sin^2 \theta - \cos^2 \theta}}
		\label{eq:a_plusminus}
	\end{equation}
	We observe that this formula breaks down if $\sin \theta = \cos \theta$ (i.e., if $\theta = \pi/4$). 
	In this case $\cos \theta = 1 /\sqrt{2}$ and from \eqref{eq:cos_theta_of_a} we obtain $a=1/2$, which corresponds to the CV whose gradient is $(\mathbf{e}_x + \mathbf{e}_y)/2$. 
	The physically acceptable solutions given by \eqref{eq:a_plusminus} are $a_+$ when $\theta \in [0, \pi/4[$  and $a_-$ when $\theta \in\ ]\pi/4, \pi/2]$

	\subsection{Determination of the \enquote{Trial} Separatrix}
	
	Corresponding to the chosen CV (i.e., to $\theta$) is the \enquote{trial} separatrix $\mathcal{S}(\theta)$, which is a line having the equation
	\begin{equation}
		y = m(x - x_\mathrm{max}) ,
	\end{equation}
	where the point $(x_\mathrm{max},0)$ is the TS. 
	The slope $m$ is determined by requiring $\nabla \xi$ to be
	orthogonal to the unit vector parallel with $\mathcal{S}(\theta)$, which is given by
	\begin{equation}
		\mathbf{e}_{\mathcal{S}(\theta)} = \frac{-\mathbf{e}_x + m \mathbf{e}_y}{\sqrt{1 + m^2}} \ .
		\label{eq:unitvec_separatrix}
	\end{equation}
	The orthogonality condition
	\begin{equation}
		\nabla \xi \cdot \mathbf{e}_{\mathcal{S}(\theta)} = (a\mathbf{e}_x + (1-a)\mathbf{e}_y) \cdot \frac{-\mathbf{e}_x + m \mathbf{e}_y}{\sqrt{1 + m^2}}  = 0
	\end{equation}
	yields $m=a/(1-a)$, which, when substituted back into \eqref{eq:unitvec_separatrix}, gives
	\begin{equation}
		\mathbf{e}_{\mathcal{S}(\theta)} = \frac{(a-1)\mathbf{e}_x + a\mathbf{e}_y}{\sqrt{a^2 + (1-a)^2}} \ .
	\end{equation}
	
	\subsection{Computation of the Probability Density}
	
	The marginal probability density is given by
	\begin{equation}
		\rho(z) = Z^{-1} \int \mathrm{d}x\ \int \mathrm{d}y\ e^{-\beta U(x,y)}\ \delta(\xi(x,y) - z)
		\label{eq:rho_seciv}
	\end{equation}
	where
	\begin{equation}
		Z = \int \mathrm{d}x\ \int \mathrm{d}y\ e^{-\beta U(x,y)} \ .
	\end{equation}
	To facilitate the evaluation of the double integrals, we transform from Cartesian coordinates to the orthogonal coordinates defined by
	\begin{eqnarray}
		q_1 = &  \xi(x,y)  =  & ax + (1-a)y \\
		q_2 = & & (a-1)x + ay \ .
	\end{eqnarray}
	That $\nabla q_1 \cdot \nabla q_2 = 0$ is manifest.
	The inverse transformation is
	\begin{align}
		x & = \frac{aq_1 + (a-1)q_2}{d} \nonumber \\
		y & = \frac{(1-a)q_1 + aq_2}{d} \ ,
		\label{eq:cartesian_of_qs}
	\end{align}
	where $d=a^2 + (1-a)^2$. 
	Hence, the Jacobian is
	\begin{align}
		\mathbf{J} & = \left( \begin{array}{cc}
			\partial x / \partial q_1 & \partial x / \partial q_2 \\
			\partial y / \partial q_1 & \partial y / \partial q_2
		\end{array} \right) \nonumber \\
		& = \left( \begin{array}{cc}
			a/d & (a-1)/d \\
			(1-a)/d & a/d
		\end{array} \right)
		\label{eq:Jacobian}
	\end{align}
	From \eqref{eq:rho_seciv} we have
	\begin{align}
		\rho(z) & = Z^{-1} \int \mathrm{d}q_1\ \int \mathrm{d}q_2\ |\mathbf{J}|\ e^{-\beta U(x,y)}\ \delta(q_1-z) \nonumber \\
		& =  Z^{-1}\int \mathrm{d}q_2\ |\mathbf{J}|\ e^{-\beta U(x,y)} \ ,
	\end{align}
	where the Cartesian coordinates that are the arguments of the PES are given in terms of $q_1=\xi$
	and $q_2$ by \eqref{eq:cartesian_of_qs}. 
	
	Using \eqref{eq:Jacobian} and the definition of $d$, we get $|\mathbf{J}|=1/d$.
	Hence, 
	\begin{equation}
		\rho(z) = \frac{\int \mathrm{d}q_2\  e^{-\beta U(x,y)}}{\int \mathrm{d}q_1\ \int \mathrm{d}q_2\  e^{-\beta U(x,y)}}
	\end{equation}

	\section{Computation of the Frequency of Crossing the Dividing Surface}
	
	We describe here the numerical implementation of the expression for $\nu$ (\eqref{eq:nu_heaviside}) in the MD simulation.

	\subsection{Details of the MD Simulation}
	
	For each simulation corresponding to a given temperature and particle mass, ten independent Langevin dynamics simulations were carried out with a friction constant of 1~ps$^{-1}$, a time step of 1~fs, and a total time of 10~ns.
	The system was propagated using the velocity Verlet algorithm.
	Crossing frequencies and activation free energies were computed for each simulation independently, only the final values were used for averages and estimation of the standard deviation.
	
	\subsection{Use of the Heaviside Function}
	
	Approximating the time derivative by the forward finite-difference formula, we rewrite \eqref{eq:nu_heaviside} as
	\begin{align}
		\nu & = \frac{1}{\tau} \sum_{i=1}^{N_f-1} \Delta t\ \left| \frac{\Theta(\xi(t_{i+1}) - z_\mathrm{TS}) - \Theta(\xi(t_{i}) - z_\mathrm{TS}) }{\Delta t} \right| \nonumber \\
		& = \frac{1}{\tau} \sum_{i=1}^{N_f-1}  \left| {\Theta(\xi(t_{i+1}) - z_\mathrm{TS}) - \Theta(\xi(t_{i}) - z_\mathrm{TS}) }\right| \ ,
		\label{eq:nu_implementation1}
	\end{align}
	where the summation on $i$ is over consecutive time steps of length $\Delta t$, $N_f$ is the number of steps of the MD simulation,  and $\tau=(N_f-1)\Delta t$ is the duration of the MD trajectory.
	That the expression in \eqref{eq:nu_implementation1}, which is straightforward to implement, yields a proper count may be seen as follows.
	If, at time step $i$, $\xi(t_i)$ and $\xi(t_{i-1})$ are both greater than or less than $z_\mathrm{TS}$, the contribution is zero, since $z_\mathrm{TS}$ is not crossed during the step. 
	If, on the other hand, $\xi(t_{i-1}) < z_\mathrm{TS}$ and $\xi(t_{i}) > z_\mathrm{TS}$, or $\xi(t_{i-1}) > z_\mathrm{TS}$ and $\xi(t_{i}) < z_\mathrm{TS}$, then the contribution is 1, as $z_\mathrm{TS}$ is crossed in one direction or the other during the step. 
	We note that $\Delta t$ must be sufficiently small that highly frequent crossings are not inadvertently missed.
	
	\subsection{Use of the Dirac Delta Function and the Atomic Velocities}
	
	We consider here an alternative approach to the computation of $\nu$.
	We begin by recasting \eqref{eq:nu_prod_absolutes} as
	\begin{equation}
		\nu = \lim\limits_{\tau \rightarrow \infty} \frac{1}{\tau} \int_0^\tau \mathrm{d}t\ \delta[\xi(t) -z_\mathrm{TS}] \left|(\nabla_x \xi)^\mathrm{T} \cdot \dot{\mathbf{x}}(t) \right| \ .
		\label{eq:nu_recast}
	\end{equation}
	Here we express the rate of change of the CV as
	\begin{equation}
		\dot{\xi}(t) = \frac{\mathrm{d}\xi}{\mathrm{d}\mathbf{x}} \cdot \frac{\mathrm{d}\mathbf{x}}{\mathrm{d}t} = (\nabla_x \xi)^\mathrm{T} \cdot \dot{\mathbf{x}}(t) \ ,
	\end{equation}
	where $(\nabla_x \xi)^\mathrm{T} = \left( \partial \xi / \partial x_1, \partial \xi / \partial x_2, \dots, \partial \xi / \partial x_{3N}  \right)$ is the 3N-dimensional gradient. 
	Discretizing the integration on time, we rewrite \eqref{eq:nu_recast} as
	\begin{align}
		\nu & = \frac{1}{\tau} \sum_{i=0}^{N_f -1} \Delta t\ \delta[\xi(t_i) -z_\mathrm{TS}] \left|(\nabla_x \xi(t_i))^\mathrm{T} \cdot \dot{\mathbf{x}}(t_i) \right| \nonumber \\
		& = \frac{1}{N_f -1} \sum_{i=0}^{N_f -1}  \delta[\xi(t_i) -z_\mathrm{TS}] \left|(\nabla_x \xi(t_i))^\mathrm{T} \cdot \dot{\mathbf{x}}(t_i) \right| \ .
		\label{eq:nu_discretized}
	\end{align}
	In practice, of course, the duration of the MD simulation, and therefore the number of time steps, are finite. 
	It is very unlikely that during a finite simulation $\xi(t_i)$ is ever exactly equal to $z_\mathrm{TS}$. 
	Hence, almost every configuration $\mathbf{x}(t)$ gives zero contribution. 
	To circumvent this problem we introduce a continuous function to represent the delta function approximately. 
	We begin by defining the continuous approximation to the Heaviside function
	\begin{equation}
		\Theta(x; \alpha) = \frac{1}{1 + e^{-\alpha x}} \ ,
	\end{equation}
	where $\alpha$ is a positive real number having dimension reciprocal length. 
	(Observe that we can formally express the true Heaviside \enquote{function} by $\Theta(x) = \lim\limits_{\alpha \rightarrow \infty} \Theta(x; \alpha)$.) 
	The corresponding delta function is given by the derivative
	\begin{equation}
		\delta(x; \alpha) = \frac{\mathrm{d} \Theta(x; \alpha)}{\mathrm{d}x} = \frac{ \alpha\ e^{-\alpha x}}{(1 + e^{-\alpha x})^{2}} \ .
		\label{eq:delta_alpha}
	\end{equation}
	Note that $\delta(x; \alpha)$ satisfies exactly the relation
	\begin{equation}
		\int_{-\infty}^\infty \mathrm{d}x\ \delta(x; \alpha) = 1 \ .
	\end{equation}
	We can now rewrite \eqref{eq:nu_discretized} as
	\begin{equation}
		\nu_\alpha = \frac{1}{N_f -1} \sum_{i=0}^{N_f -1}  \delta[\xi(t_i) -z_\mathrm{TS}; \alpha] \left|(\nabla_x \xi(t_i))^\mathrm{T} \cdot \dot{\mathbf{x}}(t_i) \right| \ ,
		\label{eq:nu_implementation2}
	\end{equation}
	where $\delta[\xi(t_i) -z_\mathrm{TS}; \alpha]$ is approximated by \eqref{eq:delta_alpha}. 
	The index $\alpha$ emphasizes the dependence of the crossing frequency on this methodological parameter.

	\begin{figure*}[h]
		\centering
		\includegraphics[height=0.9\textheight]{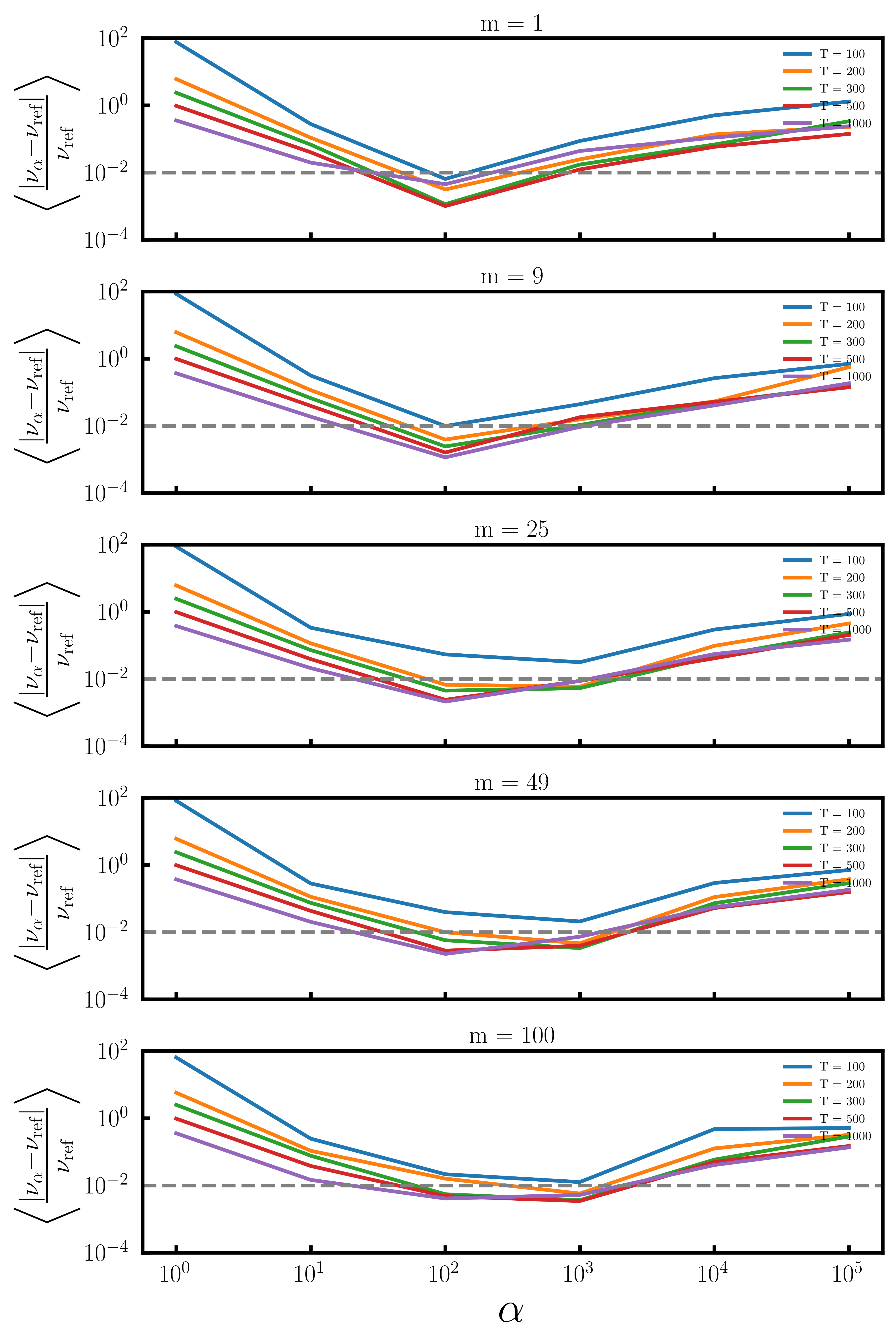}
		\caption{Plots of the magnitude of relative difference between reference crossing frequency computed by eq.~(S35) and that computed by eq.~(S40) versus $\alpha$ for the one-dimensional model PES in Section~V~A ($\epsilon = 5$~kJ/mol) for a selection of masses and temperatures. 
			Each curve is an average over 10 independent MD trajectories.
			The grey dashed line marks the threshold of 1~\% relative deviation. }
		\label{fig:nu_of_alpha}
	\end{figure*}

	The quality of this approximation depends on the choice of $\alpha$. 
	On one hand, if $\alpha$ is too small, MD frames that are far away from the dividing surface contribute significantly, thus yielding too large $\nu$. 
	On the other hand, if $\alpha$ is too large, all frames are weighted so lightly that $\nu$ is too small.
	The influence of $\alpha$ is shown graphically in Fig.~\ref{fig:nu_of_alpha}.
	Note that we plot the magnitude of the difference so that small differences are visible on the logarithmic scale.
	One can clearly see that extreme choices of $\alpha$ can lead to errors in $\nu$ of up to a factor of 100. 
	As predicted, $\nu_\alpha$ tends to zero for very large $\alpha$, which is reflected by the curves approaching 1 ($10^0$).
	The optimal choice of $\alpha$ seems to be between $10^2$ and $10^3$.
	Even choices in the range $10^1$ to $10^4$ generally yield $\nu_\alpha$'s which are around 0.9$\nu_\mathrm{ref}$ to 1.1$\nu_\mathrm{ref}$, which corresponds at 300~K to an error of approximately 0.25~kJ/mol in $\Delta F_\mathrm{RP}^\ddagger$.  
	\\

	\section{Plots of the FEPs for Models of Section~V~A with Large $\epsilon$}
	
\begin{figure}[!htb]
		\centering
		\includegraphics[width=0.66\textwidth]{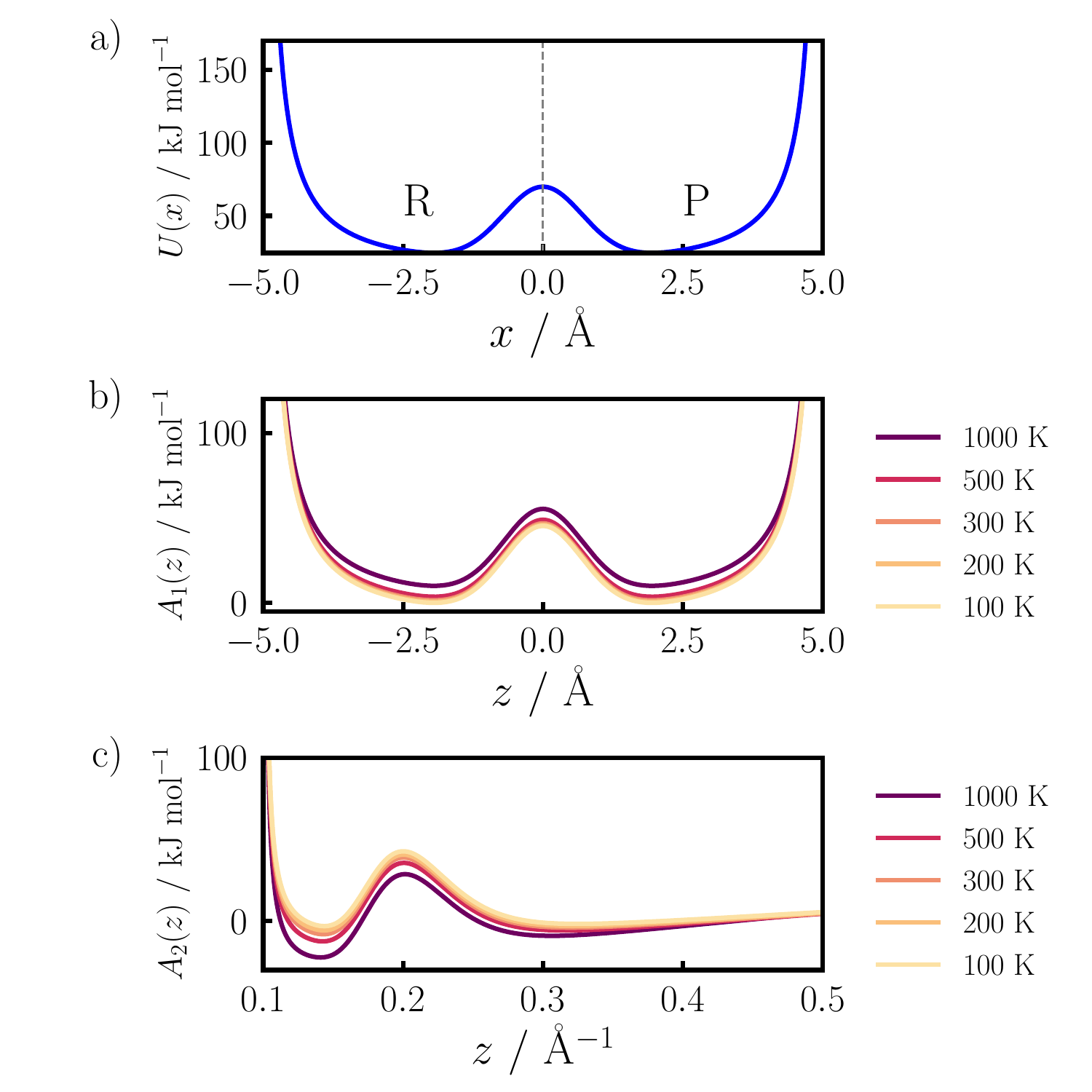}
		\caption{a) PES $U(x)$ with $\epsilon = 50$~kJ/mol (\eqref{eq:U1}); b) FEP for CV $\xi = x$ ; c) FEP for CV $\xi = \frac{1}{x+5}$.}
		\label{fig:U1b}
	\end{figure}

	In Section~V~A of the article we consider two CVs: $\xi_1(x) = x$ and $\xi_2(x)=1/(x+5)$.
	Fig.~\ref{fig:U1b}a shows the the PES \eqref{eq:U1}	for the case $a=1$~\AA$^{-2}$, $b=1$~\AA, and $\epsilon = 50$~kJ/mol.
	Figs.~\ref{fig:U1b}b and~\ref{fig:U1b}c show plots of the FEPs based on the two CVs just as does Fig.~5 of the article. 
	
	\section{Computation of FEPs for the Cyclization of the Hexenyl Radical}
	
	\textit{Ab-initio} MD simulations on DFT level were performed using a development version of the FermiONs++ program package \cite{kussmann2013pre,kussmann2015preselective,laqua2020highly,laqua2021accelerating}. 
	For this purpose the $\omega$B97M-V functional was applied with the def2-TZVP basis set \cite{mardirossian2016omega, schafer1992fully}. 
	To account for solvation in benzene the COSMO continuum solvation model was used \cite{klamt1993cosmo}.
	An optimized minimum energy structure was heated from 0.1~K to 310~K over 3100 time steps with a step size of 0.1~fs.
	Initial momenta were randomly drawn from the Maxwell-Boltzmann distribution. 
	Velocities were re-scaled every 10 time steps to increase the temperature by 1~K. 
	For production runs the temperature was controlled by a Langevin thermostat with friction coefficient 0.001~fs$^{-1}$ at 300~K. The time step was set to 0.5~fs. 
	The dynamics was biased along reaction coordinates $\xi = d(\mathrm{C1-C5})$ 
	with the WTM-eABF method \cite{fu2018zooming,fu2019taming} applying a recently published Python implementation \cite{Hulm2022}.
	For simulations along $\xi$ the extended-variable was coupled to the reaction coordinate with a thermal width of 0.05~\AA\ and the system was confined with harmonic walls at 1.0~\AA\ and 6.0~\AA. 
	The bias force was stored on a grid with bin width 0.05~\AA. 
	The ABF force was scaled up linearly and the full bias applied in bins with more that 200 samples. 
	For the Well-Tempered Metadynamics (WTM) potential Gaussian kernels of height 0.5~kJ/mol and standard deviation 0.1~\AA\ for $\xi$ were deposited every 20~steps. 
	The height of new Gaussian hills was scaled down over the course of the simulation with effective temperature of 2000~K. 
	Sampling of $\xi$ was performed with a single walker running for about 290~ps.
	
	To obtain thermodynamic properties statistical weights of individual frames were recovered in post-processing using the MBAR algorithm \cite{Shirts2008}.
	For this purpose the sampled probability distribution was repartitioned into a mixture of Gaussian distributions with standard deviation 0.05~\AA\ for $\xi$
	\cite{Hulm2022}.

\end{document}